\documentclass[aps,prb,twocolumn,showpacs,superscriptaddress]{revtex4-1}
\usepackage{amsmath} 
\usepackage{graphicx} 
\usepackage{hyperref}

\newcommand{\e}[1]{\mathrm{e}^{#1}}

\begin{document}
\title{Discontinuous polaron transition in a two-band model}

\author{Mirko M. M\" oller} \affiliation{\!Department \!of \!Physics and
  Astronomy, \!University of\! British Columbia, \!Vancouver, British
  \!Columbia,\! Canada,\! V6T \!1Z1} 
\author{Mona Berciu} \affiliation{\!Department \!of
  \!Physics and Astronomy, \!University of\! British Columbia,
  \!Vancouver, British \!Columbia,\! Canada,\! V6T \!1Z1}
\affiliation{\!Quantum Matter \!Institute, \!University of British
  Columbia, \!Vancouver, British \!Columbia, \!Canada, \!V6T \!1Z4}

\begin{abstract}
We present exact diagonalization and momentum average approximation
(MA) results for the single polaron properties of a one-dimensional
two-band model with phonon-modulated hopping. At strong
electron-phonon coupling, we find a novel type of sharp transition,
where the polaron ground state momentum jumps discontinuously from
$k=\pi$ to $k=0$. The nature and origin of this transition is
investigated and compared to that of the Su-Schrieffer-Heeger (SSH)
model, where a sharp but smooth transition was previously reported. We
argue that such discontinuous transitions are a consequence of the
multi-band nature of the model, and are unlikely to be observed in
one-band models. We also show that MA describes qualitatively and even
quantitatively accurately this polaron and its transition. Given its
computationally efficient generalization to higher dimensions, MA thus
promises to allow for accurate studies of electron-phonon coupling in
multi-band models in higher dimensions.
\end{abstract}

\maketitle
    
\section{Introduction}
The coupling between carriers and phonons is known or believed
to be important for many materials, including cuprates,\cite{cuprates,
Jahn-Teller-Polaron, manganites+cuprates,cuprates-phonons-review}
manganites, \cite{Jahn-Teller-Polaron, manganites+cuprates,
manganites-phonons}, nickelates \cite{Steve-nickelates,
Jahn-Teller-Polaron, nickelates-normal-modes, nickelates-Zaanen} and
bismuth perovskites.\cite{Kateryna,
Franchini-BaBiO3} These materials display a variety of interesting
phenomena, including, but not limited to, high-temperature
superconductivity (cuprates, BaBiO$_3$), layered ferromagnetism
(manganites) and a spin/charge density wave (nickelates).

The carrier-phonon coupling leads to the formation of a polaron, a
coherent quasi-particle (QP) consisting of the charge carrier and the
cloud of phonons surrounding it and moving coherently with it.
Polarons have been studied extensively especially in the Holstein
model, \cite{Bonca+Trugman1, Bonca+Trugman2, Bonca+Trugman3, Li,
  Alvermann+Trugman, Holstein-Long-Range, Mona-MA-1, Mona-MA-2} the
simplest model where local phonons modify  the on-site energy of
the carrier, but also to generalizations with short-range and
long-range couplings of similar origin, such as the breathing-mode (BM) model
\cite{Bayo, CBM, Glen-BM}, the double-well potential
model\cite{Clemens1, Clemens2} and the Fr\"ohlich model \cite{Froehlich1, Froehlich2}.

The other possibility is that the coupling to phonons modulates the
carrier's hopping integrals, a scenario described by the SSH model,
\cite{SSH-original-paper} which has seen an increased amount of
interest in recent years.\cite{DominicPRL,Zhao,MonaPRL110} This is
because the SSH model exhibits a sharp transition in the properties of
its polaron, one signature being the change of the polaron ground
state (GS) momentum from $k=0$ (at weak coupling) to a finite value
that smoothly evolves toward $k=\pi/2$ (at strong coupling). Such
transitions were shown to be impossible for models where the phonons
modulate the on-site energy \cite{Gerlach}.A study of a model which
includes both types of carrier-phonon coupling was carried out by
Herrera {\em et. al} \cite{MonaPRL110} and found that in addition to
the transition observed in the SSH model, a second transition of the
GS momentum also takes place. Whether other such transitions can occur
and what are their characteristics, is  currently an open
question.
 
\begin{figure}[b]
  \centering \includegraphics[width=\columnwidth]{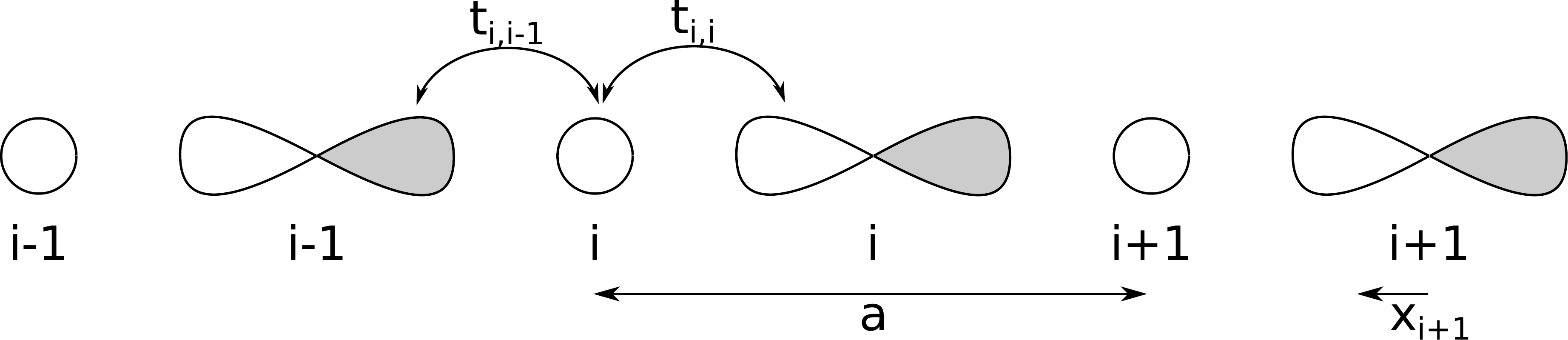}
  \caption{Sketch of the model, including the convention we use for
    indexing the sites and for the orientation of the lobes. The sign convention for the displacement $x_{i+1}$ due to a phonon on site $i+1$ is also shown. }
  \label{fig:model}
\end{figure}

Efforts to understand polaron physics have, so far, focused almost
exclusively on single-band models. It is therefore a natural question
to ask whether the polaron properties of multi-band models are
similar, or whether they are qualitatively different from those of
single band models. In this Article, we answer this question based on
a study of the single polaron properties of the two-band model
depicted in Fig.  \ref{fig:model}, with two different atoms per unit
cell, one of which is light and thus supports lattice vibrations
(optical phonons). The coupling to this phonon mode modulates the
hopping integrals, in direct analogy with the one-band SSH model.(For
the vanishing carrier concentration of interest to us, the SSH-model
is a one-band model because the Peierls dimerization only occurs at
half-filling \cite{DominicPRL}).

Given this similarity, one may expect our two-band model to behave
like the SSH model, and indeed we find a sharp transition at strong
carrier-hole coupling, where the polaron GS momentum changes its
value. However, unlike in the one-band SSH model where the GS momentum
changes continuously with coupling, in our model the GS momentum jumps
discontinuously from $k=\pi/a$ to $k=0$. Furthermore, this transition
leads to an extreme flattening of the polaron band, unlike in the SSH
model. We conclude that this transition is qualitatively different
from that of the one-band SSH model. We argue that the mechanism for
the transition is a competition between the bare carrier hopping which
favors a GS momentum $k=\pi/a$, and the phonon-modulated hopping which
favors $k=0$. While some of this physics is similar to that explaining
the transition in the SSH model, we find strong indications that the
two-band nature of our model plays a vital role. This makes it
unlikely that such a discontinuous transition can occur in a one-band
model. The physics described in this article is therefore, to our
knowledge, fundamentally new and our findings hint at the possibility
that there is still more, new polaron physics to be discovered in
other multi-band models.

The physical inspiration for our model is the perovskite BaBiO$_3$,
which is known to have strong electron-phonon coupling and to exhibit
superconductivity up to surprisingly high temperatures upon hole
doping ($T_C \sim 30K$ in the case of Ba$_{1-x}$K$_x$BiO$_3$
\cite{BaKBiO3-Tc} and $T_C \sim 13K$ for BaPb$_{1-x}$Bi$_{x}$O$_3$)
\cite{BaPbBiO3-Tc}, widely believed to be due to a phononic glue
\cite{Kotliar-PRL, Kotliar-PRX, EPC-tilts}. The relevant valence
orbitals are the Bi 6$s$ orbitals and the O 2$p$ orbitals.

Previous work on polarons and bipolarons in this material was carried
out by Allen {\em et al.} \cite{Allen-half-filling,
Allen-single-carrier} and is based on the Rice-Sneddon model
\cite{Rice-charge-disp} which assumes that the Bi atoms undergo charge
disproportionation, Bi$^{4+}$Bi$^{4+}\rightarrow
$Bi$^{3+}$Bi$^{5+}$. This scenario has received wide-spread attention
\cite{Cox-charge-disp-1, Cox-charge-disp-2, Rice-charge-disp,
Varma-charge-disp, Hase-charge-disp, Sleight-review, Franchini-BaBiO3,
Allen-half-filling, Allen-single-carrier}, and as a consequence many
model Hamiltonians only take into account the Bi 6$s$
orbitals. Polaronic signatures in agreement with this work have also
been found experimentally \cite{Nishio,Guo-meng-Zhao}.

A different scenario is provided by a recent study of Foyevtsova {\em
et. al.} \cite{Kateryna} and the experimental as well as theoretical
work of Menushenkov {\em et al} \cite{Menushenkov-1,
Menushenkov-2}. Foyevtsova {\em et. al.} argue that the BiO$_6$
octahedra undergo a breathing distortion due to strong hybridization
between the Bi 6$s$ and O 2$p$ orbitals, with the holes being located
primarily on the O. This picture is similar to that proposed recently
for the nickelates \cite{George-nickelates, Millis1, Millis2,
Steve-nickelates} and the inspiration for our ``toy model''. In this
picture both the Bi 6$s$ and O 2$p$ orbitals need to be taken into
account. Since O atoms are much lighter than Bi atoms, only optical
phonons on the O atoms are considered and are allowed to modulate the
hopping integrals.

When compared to other perovskites such as the cuprates, manganites
and nickelates, BaBiO$_3$ is appealing because of its comparatively
simple electronic structure and absence of magnetic properties.  To
simplify things even more, instead of considering a model describing
such a material at or near half filling, as it is in reality, we
investigate the single polaron physics in its almost fully compensated
case, {\em i.e.} like in LaBiO$_3$ with one extra hole. Generalizing
to one carrier (polaron) per unit cell will be left for future
work. Moreover, as indicated above, we restrict ourselves to study a
1D BiO-like chain, instead of treating the full 3D system. There are
two practical reasons for these simplifications: (i) they make
comparison to the SSH model, where polaron results are currently
available only for the 1D model, possible; (ii) they allow us to use
exact diagonalization (ED) to find essentially exact results very
efficiently,\cite{Bonca+Trugman1} which in turn also allows us to
probe a wide range of coupling strengths and phonon frequencies to
understand the relevant physics.

Beside using ED to understand the polaron properties of this model, we
also develop and validate here two simple versions of the variational
momentum average (MA) approximation,\cite{Mona-MA-1, Mona-MA-2,
  Glen-BM} which capture the relevant polaronic physics qualitatively
and even quantitatively. MA approaches are very useful because their
accuracy improves in higher dimensions while maintaining similar
computational efficiency. This is in contrast to ED and most other
numerical methods that become very costly due to the significant
increase of the Hilbert space in higher dimensions. This work can
therefore be seen as a first step towards a study of the 3D
systems. Note that apart from their usefulness in treating more complex
problems, developing such approximations also leads to a better
understanding of the nature of the polaron's cloud.

To summarize, the research presented in this article serves two main
purposes: (i) to reveal surprising, new polaron physics whose nature
appears to be tied to the two-band nature of our model, and (ii) to
serve as a test-ground for approximations which will be useful in
solving the 3D many-band problem, and thus help to improve our
understanding of the fascinating compound BaBiO$_3$, if the scenario
of Foyevtsova {\em et. al.} turns out to be valid, or of other
many-band materials with hopping-modulated carrier-phonon coupling.

The remainder of this article is organized as follows: in Sec.
\ref{sec:model} we introduce the model and in Sec. \ref{sec:methods}
we introduce the ED algorithm, MA and a perturbation theory.  The
results are presented in Sec. \ref{sec:results} and Sec.
\ref{sec:conclusions} contains our conclusions.

\section{Model}
\label{sec:model}

We study the single polaron properties of the 1D, two-band model sketched
in Fig. \ref{fig:model}. There are two atoms per unit cell, one
hosting  valence electrons in an $s$-orbital, and the other in a
$p$-orbital. The latter  atom is assumed to be sufficiently light so
that it is a good approximation to ignore the motion of the heavier
ones. In other words, we assume that the lighter atoms oscillate inside the ``cage''
made of heavier atoms, giving rise to an optical phonon mode that
modulates the $s$-$p$ hopping of the carriers. We consider the
limit of a very lightly doped insulator, {\em i.e.} all states are
filled except for a single hole present on the chain. The inspiration
to study such a  model was discussed in the previous section.

The kinetic energy of the hole is described by a nearest neighbor
tight-binding Hamiltonian:
\begin{align}
  \hat{T}_{\rm tot} = \sum_{i} (t_{i,i-1}s_{i}^\dagger p_{i-1} +
  t_{i,i} s_{i}^\dagger p_i +\text{h.c.}).
\end{align}
Here $s_i^\dagger\ (p_i^\dagger)$ creates a hole on the $s$ ($p$)
orbital of the atoms in the $i^\text{th}$ unit cell (the spin is an
irrelevant degree of freedom here, and we do not write it explicitly). Their
Fourier transforms are: $s_k^\dagger = \sum_j \exp(i k R_j)/\sqrt{N}
s_j^\dagger$ and $p_k^\dagger = \sum_j \exp(i k R_j)/\sqrt{N}
p_j^\dagger$, where $R_j = j a$ is the location of the unit cell
$j=1,\dots, N$, and the number of unit cells $N\rightarrow \infty$ .
The momentum $k$ is restricted to the first Brillouin zone (BZ),
$-{\pi} < ka \le {\pi}$.

For an undistorted chain (no phonons), and keeping in mind that $s_i^\dagger$ and $p_i^\dagger$ are hole operators, for the choice of lobe
orientation shown in  Fig. \ref{fig:model} we have $t_{ii}=-t_{i,i-1}=t$, leading
to the kinetic energy:
\begin{align}
  \hat{T} = t\sum_{i} (-s_{i}^\dagger p_{i-1} + s_{i}^\dagger p_i
  +\text{h.c.})
\end{align}
The $t>0$ hopping parameter is given by the overlap of the $s$ and
$p$-orbitals when the atoms are in their equilibrium positions.
Changing its sign corresponds to changing the sign convention for the
$p$-orbitals, and therefore does not have any physical effect. As a
consequence, the results we present below for a hole-doped chain
remain identical for an electron-doped chain as well. When phonons are
excited, the hopping amplitudes change from this equilibrium value,
resulting in the hole-phonon coupling term discussed below.

The hole's on-site energy depends on whether it sits on an $s$- or a
$p$-orbital and leads to a charge transfer term:
\begin{align}
  \hat{H}_{\text{ct}} = -\Delta \sum_{i} p_i^\dagger p_i.
\end{align}
The difference in on-site energies can have either sign, favoring the
$p$ (if $\Delta>0$) or $s$ (if $\Delta<0$) orbitals. The on-site energy for the $s$ orbitals is set to zero.

The phonons are assumed to be described by a dispersionless Einstein mode with
energy $\Omega$ (we set $\hbar=1$):
\begin{align}
  \hat{H}_{\text{ph}} = \Omega \sum_i b_i^\dagger b_i,
\end{align}
where $b_i^\dagger$ creates a phonon on the  $p$-orbital of the lighter atom of
the $i^\text{th}$ unit cell. As discussed, the heavier atoms are taken
to be frozen in their equilibrium positions.

This model allows for two types of hole-lattice coupling. One comes from
the modulation of on-site energies, because when the distance
between neighbor atoms changes, so do the corresponding Coulomb
interactions. We are not aware of a comprehensive study of this type
of coupling for this two-band model, although the asymptotic cases
with $\Delta \rightarrow -\infty$ (the breathing-mode model
\cite{Bayo,Glen-BM}) and $\Delta \rightarrow +\infty$ (the double-well
model\cite{Clemens1, Clemens2}) have been studied and have revealed
interesting polaronic behavior, although still rather
conventional.

Instead, we focus here on the hole-lattice coupling arising from the
fact that changes in the distance between adjacent atoms also modulate
the hopping integrals. As we show below, this modulation of the
kinetic energy leads to qualitatively new physics of a kind that, so
far as we know, has not been revealed before.

Thus, the hole-phonon coupling that we study arises from the linear
expansion of the hopping amplitudes $t_{i,i}$ and $t_{i,i-1}$ as a
function of the small oxygen displacement $x_i \propto b_i +
b_i^\dagger$. If we choose coordinates such that $x_i>0$ for a
displacement toward the left, to linear order this expansion gives:
\begin{align}
  &t_{i,i} \approx t[1+\alpha x_{i}] = t[1+\tilde{\alpha} (b_i +
    b_i^\dagger)] \\ &t_{i,i-1} \approx -t[1-\alpha x_{i-1}]
 = -t[1-\tilde{\alpha} (b_{i-1} + b_{i-1}^\dagger)]
\end{align}
Within this approximation $\hat{T}_{\rm tot} =\hat{T}+
\hat{H}_{\text{h-ph}} $, where after absorbing all the constants into
the coupling $g$, we find:
\begin{align}
  \hat{H}_{\text{h-ph}}= g \sum_i\left[ (s_i^\dagger p_i +
    s_{i+1}^\dagger p_i) (b_i + b_i^\dagger) + \text{h.c.}\right]
\end{align}
We choose $g>0$. Note that because the sign of $g$ is controlled by
the choice of the coordinate system, a change $g \rightarrow -g$ is
equivalent with choosing $x_i>0$ for displacements toward the right.
Consequently the polaron properties only depend on the magnitude of
$g$, not on its sign.

\begin{figure}[t]
  \centering \includegraphics[width=\columnwidth]{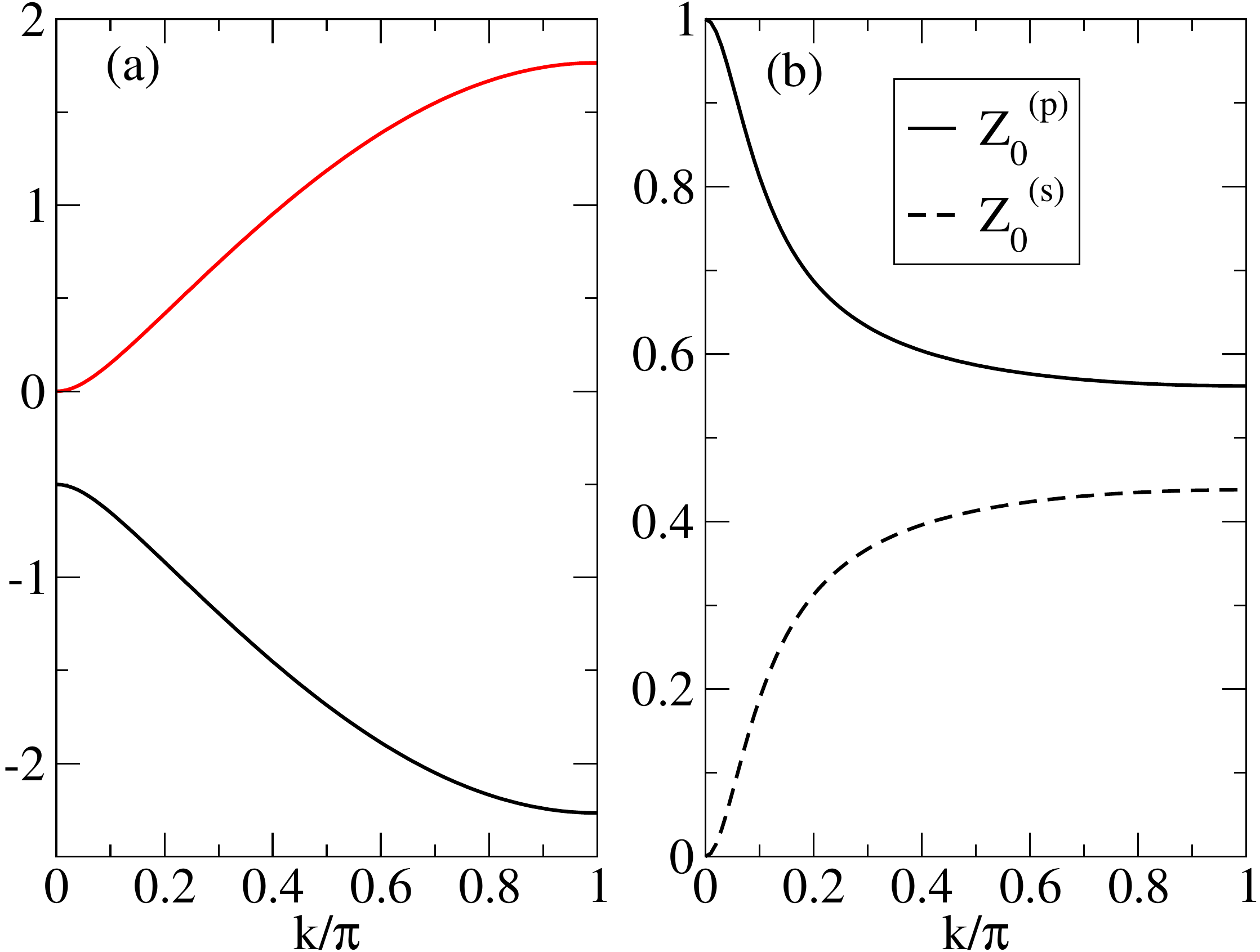}
  \caption{(color online) (a) Free hole band structure. (b) The free-hole
    weights $Z_{0}^{(s)}(k)$ and $Z_{0}^{(s)}(k)$ for the lower band
    of energy $E_0(k)$. In both panels $\Delta=0.5$, $t=1$ and $a=1$.}
  \label{fig:H0}
\end{figure}

The Hamiltonian studied here is, therefore,
$$ 
\hat{H} = \hat{T} + \hat{H}_{\text{ct}} + \hat{H}_{\text{ph}} +
\hat{H}_{\text{h-ph}}.
$$ 
Before moving on, we briefly comment on its main limitations.  Keeping
only linear terms in the expansion of the hopping integrals is a valid
approximation when $g$ is sufficiently small, so that the polaron
cloud creates rather small local distortions. For large values of $g$,
the effects of higher order coupling terms $\propto
(b_i+b_i^\dagger)^n$ with $n>1$, need to be considered because now the
local distortions become large so the displacements $x_i$ may no
longer be assumed to be small. This is true for all models with
electron-phonon coupling. For the Holstein and double-well potential
models, the effect of such non-linear terms was shown to be
significant in the strong coupling limit (as defined by the linear term).\cite{Clemens1, Clemens2, Clemens3}

Another approximation is to consider only nearest neighbor (nn)
hopping. Coming back to the inspiration for this model, for a
BiO-like chain one could argue that the Bi 6$s$ orbitals are quite
broad and therefore next nearest neighbor (nnn) hopping between them
may play a role. Furthermore, in  3D models  $t_{pp}$ hopping between nn
O 2$p$ orbitals needs to be included. In 1D, however, this type of
hopping is much less relevant because O atoms are separated by at least
on Bi atom. 

Finally, as already mentioned, hole-phonon coupling (linear or to
higher order) resulting from the modulation of the on-site energies
could also be included in this model. Nevertheless, here we study
polaron properties for the linear model discussed above, for a wide
range of $g$ values, to explore its physics and to provide a baseline
from which the effect of such additional terms can be gauged. We also
note that all such further extensions can be studied with the methods
we use in this work.

The free-hole dispersion is obtained by diagonalizing $\hat{H}_0 =
\hat{T} +\hat{H}_{\text{ct}} $. This results in two bands
with eigenenergies $ -\frac{\Delta}{2} \pm \sqrt{\frac{\Delta^2}{4} +
  4 t^2 \sin^2{ka \over 2}}$, shown in Fig. \ref{fig:H0}(a). The lower
band, which will be referred to as $E_{0}(k)$, has its minimum at
$k=\pi/a$ where the hybridization between $s$ and $p$-orbitals is
maximal. In contrast, the hybridization vanishes at $k=0$.

The free-hole weights $Z^{(s,p)}_{0}(k)$ are defined as
\begin{align}
  &Z^{(s)}_{0}(k) = |\langle \Phi_0(k)| s_k^\dagger |
  0\rangle |^2 ; &Z^{(p)}_{0}(k) = |\langle \Phi_0(k)|
  p_k^\dagger | 0\rangle |^2
\end{align}
and measure the overlap between the low-energy eigenstate
$|\Phi_0(k)\rangle$ and the free hole in a pure $s-$ or
$p$-state, respectively, and are given by:
\begin{align}
  Z^{(s)}_{0}(k) = 1-Z^{(p)}_{0}(k) = \frac{4t^2 \sin^2(ka/2)} {4t^2
    \sin^2(ka/2) + E_0^2(k)}
\end{align}
Note that $Z^{(s)}_0(0) = 0$ confirms that for $k=0$, the low-energy
free hole sits only on the $p$-orbitals.

We are interested in the evolution of this low-energy band as the
coupling to phonons is turned on, and in the nature of the resulting
quasiparticle -- the polaron.

\section{Methods}
\label{sec:methods}
\subsection{Perturbation Theory}
\label{sec:perturbation-theory}

A lot of insight can be gained from studying the anti-adiabatic limit
$\Omega \gg t$, where at sufficiently small $g$ the phonon
cloud of the polaron is very small, {\em i.e.} with at most one phonon
present. The energy correction must be of order $g^2$ because
$H_{\text{h-ph}}$ changes the phonon number and thus has vanishing
average value in the free hole ground state.

Consider the effects of $H_{\text{h-ph}}$. For an $s$-orbital hole, it
allows it to hop onto the adjacent $p$-orbital while also emitting a
phonon at this $p$-orbital. The phonon needs to be reabsorbed, which
can occur in two ways: {(i)} by hopping to the next $s$-orbital, or
{(ii)} by hopping back to the original $s$-orbital. The first process
results in an effective $s$-$s$ hopping of amplitude $-2g^2/\Omega
\cos(k a)$, whereas the second changes the on-site $s$ energy by $-2
g^2/\Omega$.

For a hole starting from a $p$-orbital, emission of a phonon
requires the hole to hop to an adjacent $s$-orbital. From there the
phonon can only be reabsorbed if the hole returns to the original
$p$-orbital. This process gives an additional on-site $p$ energy of
$-2 g^2/\Omega$. Thus, the effective Hamiltonian is:
\begin{align}
 &\hat{H}_{\text{eff}}  = \left (
  \begin{array}{cc}
    -2t \lambda [1+\cos(ka)] & t(1-\e{-i k a}) \\ t(1-\e{i k a}) & -2t
    \lambda - \Delta \\
  \end{array} \right )
\label{eq:Heff_matrix}
\end{align}
 where we introduced the dimensionless, effective coupling $\lambda =
 g^2/(\Omega t)$. The lowest eigenenergy, {\em i.e.} the polaron
 dispersion, is given by:
\begin{align}
  E_{\mathrm{P}}^{\text{eff}}(k) = &-t \lambda [2 + \cos(k a)]
  -\frac{\Delta}{2} \nonumber \\ &-\sqrt{\left [t \lambda
      \cos(ka)-\frac{\Delta}{2} \right]^2 +4t^2 \sin^2(\frac{ka}{2})}
  \label{eq:Eeff}
\end{align}
This gives a good approximation for the polaron band in the limit
$\Omega \gg t,g$, as discussed below.

\subsection{Exact Diagonalization (ED)}

The polaron eigenstates can also be obtained by ED.  Our
implementation is a direct extension of the method proposed in
Ref. \onlinecite{Bonca+Trugman1}, which has already been successfully
applied to a variety of polaronic models: the Holstein model in
various dimensions,\cite{Bonca+Trugman1, Bonca+Trugman3, Li,
Bonca+Trugman2, Alvermann+Trugman} a generalized Holstein model with
longer range interactions,\cite{Holstein-Long-Range} the $t$-$J$ model
with hole-phonon coupling, \cite{Bonca-t-J} and the breathing mode
model,\cite{Bayo} to name a few. We briefly review it here.

The Hilbert space is spanned by the following translationally
invariant basis states:
\begin{align}
  |\mathcal{C}, k, \sigma\rangle = \sum_i \frac{\e{i k R_i}}{\sqrt{N}}
  c_{i,\sigma}^\dagger \prod_{m \in \mathcal{C}}
  \frac{(b_{i+m}^\dagger)^{n_m}}{\sqrt{n_m!}} | 0 \rangle
\end{align}
Here $\sigma=s,p$ is an index identifying the orbital, such that
$c_{i,\text{s}}^\dagger = s_i^\dagger$ and $c_{i,\text{p}}^\dagger =
p_i^\dagger$. $\mathcal{C}$ defines specific phonon cloud
configurations, and $k$ is the total momentum.

Following Bon\v{c}a {\em et al.}, we construct the Hilbert space by
acting $M$ times with the full Hamiltonian $\hat{H}$ on the free
carrier states $s_k^\dagger |0\rangle$ and $p_k^\dagger | 0 \rangle$
and all the states which are created in this process. This quickly
generates a large enough Hilbert space to accurately calculate the
ground state energy of the polaron with the Lanczos
technique.\cite{Dagotto-Lanczos} Convergence is reached when an
increase in the value of $M$ no longer produces a change in the
eigenenergy. The number of states contained in this Hilbert subspace for
different values of $M$ is listed in Table \ref{tab:Nstates}.

\begin{table}[t]
  \centering
  \begin{tabular}{|c|c|}
    $M$ & Number of states \\
    \hline 
    10& 4 619\\
    11& 9 227 \\
    12& 18 358\\
    13& 36 314 \\
    14& 71 540 \\
    15& 140 943 \\
    16& 276 108 \\
    17& 540 923 \\
    18& 1 056 244 \\
    19& 2 062 913 \\
    20& 4 014 953 \\
  \end{tabular}
  \caption{Number of states in the Hilbert space}
  \label{tab:Nstates}
\end{table}

\subsection{Momentum average approximation (MA)}

MA is an accurate variational method for calculating propagators of
single polaron Hamiltonians, from which polaron properties such as its
energy and quasiparticle weight can be obtained. Its simplest version
was introduced for the Holstein model,\cite{Mona-MA-1,Mona-MA-3} and
then it was shown that it can be systematically improved by increasing
the size of the variational space, {\em i.e.} which phonon
configurations are included.\cite{Mona-MA-2} Apart from the Holstein
model, MA has also been shown to be very accurate for many other
lattice polaron models including the BM model,\cite{Glen-BM} the SSH
model,\cite{DominicPRL, MonaPRL110} and the double-well
model.\cite{Clemens1, Clemens2}

Here we propose two versions of MA: MA$^{(0)}$ for a variational space
containing states with a one-site phonon cloud, and MA$^{(1r)}$ which
also includes states with one additional phonon on a site adjacent to
this one-site phonon cloud. MA$^{(1r)}$ is a restricted version of
MA$^{(1)}$ which allows the additional phonon to be at any distance
from the cloud.\cite{Mona-MA-2} It can also be viewed as a simplified
version of the two-site cloud version described in Ref.
\onlinecite{Glen-BM} for the breathing mode model. The latter can be
implemented easily for this model as well but is more cumbersome to
generalize to higher dimensions. As we argue below, these simpler
versions already suffice for our purposes.

The Green's functions (GF) of interest are:
\begin{align}
  G^{\sigma'\sigma}(k,\omega) = \langle 0 | c_{k,\sigma'}
  \hat{G}(\omega) c_{k,\sigma}^\dagger | 0 \rangle,
\end{align}
where again $\sigma=s,p$ identifies the orbitals. $\hat{G}(\omega) =
[\omega - \hat{H} + i \eta]^{-1}$ is the resolvent of $\hat{H}$ and $i
\eta$ is a small positive imaginary number indicating that we are
computing retarded GFs. A complete derivation of the MA solution
can be found in Appendix A.

\section{Results}
\label{sec:results}

We first present the ED results and discuss their meaning and
implications. We then use them to gauge the accuracy of the MA$^{(0)}$
and MA$^{(1r)}$ results.

\begin{figure}[t]
  \centering \includegraphics[width=\columnwidth]{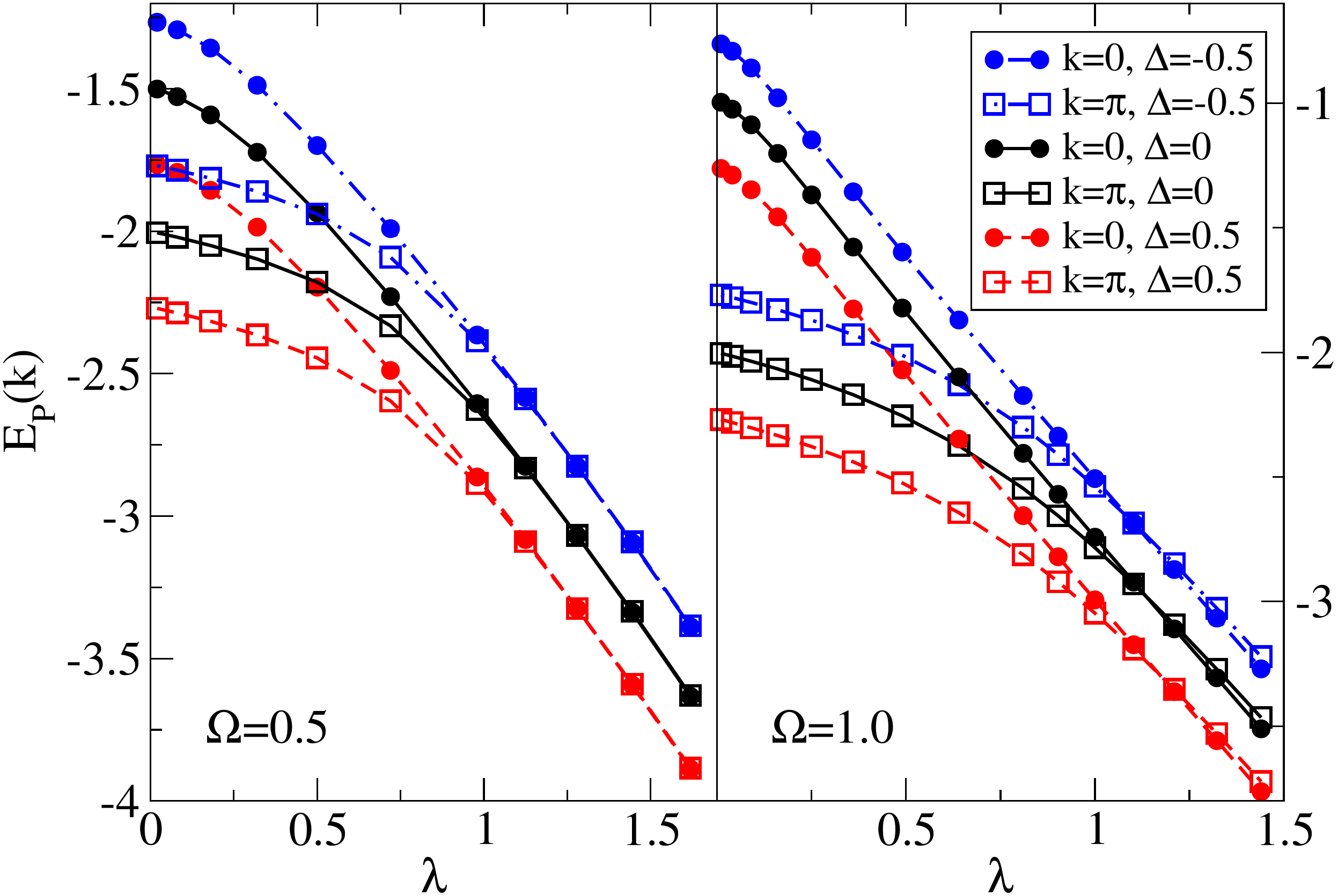}
  \caption{(color online) $E_{\text{P}}(0)$ and $E_{\text{P}}(\pi)$
    vs. the effective coupling $\lambda$ for $\Omega=0.5$ (left panel),
    and $\Omega=1.0$ (right panel), at $\Delta=0,\pm0.5$. Convergence
    was reached for $M=14,18$, respectively.}
  \label{fig:EGS_vs_g}
\end{figure}

In Fig. \ref{fig:EGS_vs_g} we plot the highest and lowest ED values of
the polaron energy, $E_{\text{P}}(0)$ and $E_{\text{P}}(\pi)$, vs. the
effective coupling $\lambda$ for two values of the phonon frequency.
Results are qualitatively similar for all other tested values of
$\Omega$. For small $\lambda$, the polaron GS energy (at $k=\pi$) is
close to the free hole energy, $E_{\text{P}}(\pi) \approx
E_{0}(\pi)=-\Delta/2-\sqrt{\Delta^2/4+4t^2}$. At $k=0$, the polaron
band lies just below the polaron+phonon continuum, so
$E_{\text{P}}(0)=E_{\text{P}}(\pi)+\Omega$. With increasing $\lambda$,
the polaron band moves to lower energies and its bandwidth narrows
considerably, as the polaron becomes heavier. All this is standard
polaronic physics.

\begin{figure}[t]
  \centering \includegraphics[width=\columnwidth]{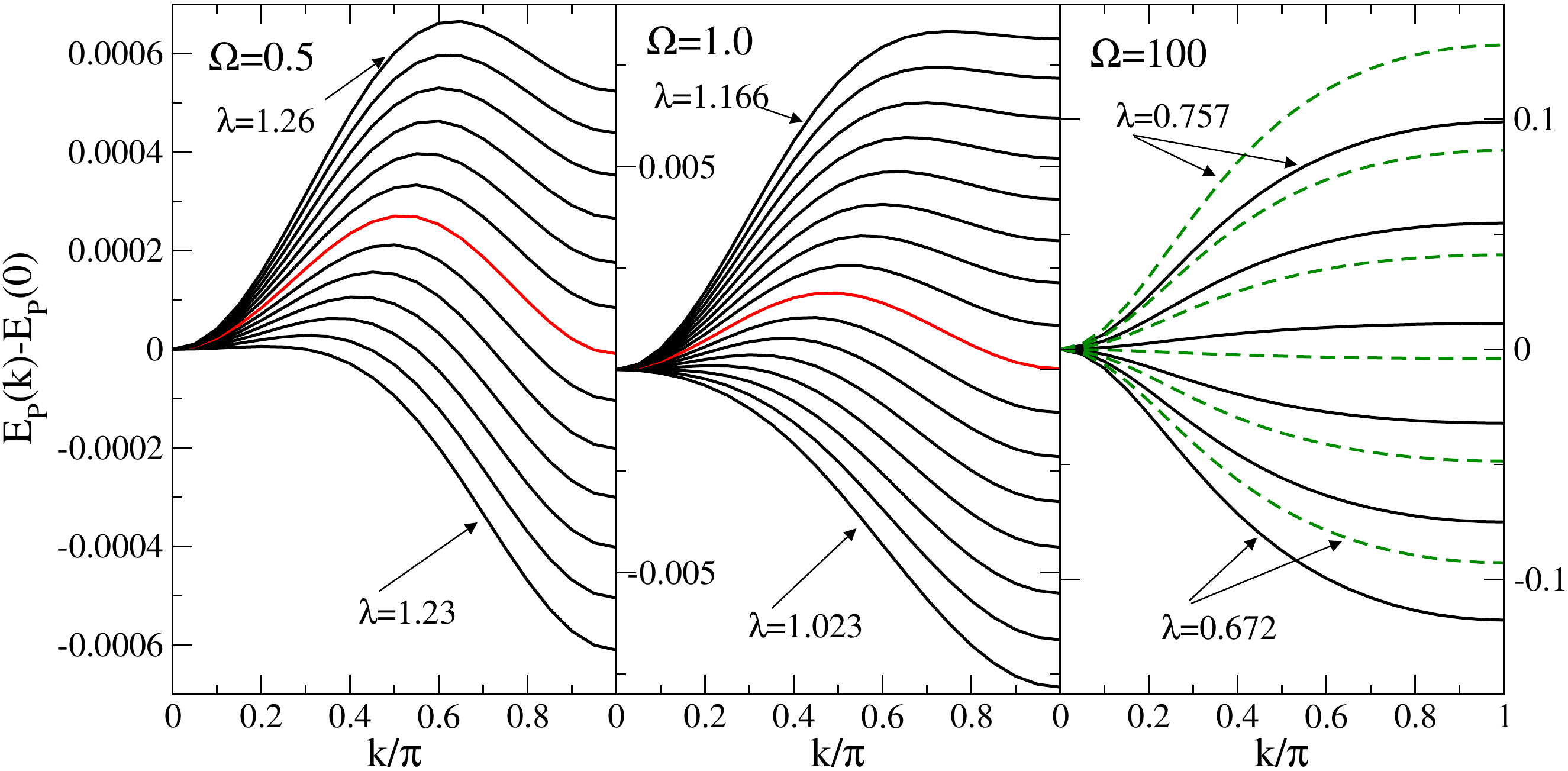} 
  \caption{(color online) Polaron dispersion for $\Omega=0.5,1.0,100$
    and values of $\lambda\sim \lambda_c$. The curves have been
    shifted by $E_{\text{P}}(0)$\ for better comparison and the curve
    at $\lambda_c$ is highlighted in red. The green, dashed lines in
    the rightmost panel are the perturbative results from Eq. 
    (\ref{eq:Eeff}). All curves are for $\Delta=0$ and the cutoff is 
    $M=10,14,18$, respectively.}
  \label{fig:dispersion_close_to_gc}
\end{figure}

The surprise is that at sufficiently
large $\lambda$, $E_{\text{P}}(0)$ and $E_{\text{P}}(\pi)$ cross,
indicating that the GS momentum changes its value. For $\Omega=0.5$,
in Fig. \ref{fig:EGS_vs_g}(a), the scale of the graph makes it
difficult to see the actual crossing, but its occurrence is verified
below, in Fig. \ref{fig:dispersion_close_to_gc}. It is not a priori
obvious that the new GS momentum is necessarily at $k=0$, but we will
show below that this is the case. Consequently, we define the critical coupling
$\lambda_c$ (or $g_c$) as the value for which
$E_{\text{P}}(\pi)=E_{\text{P}}(0)$.

\begin{figure}[t]
  \centering \includegraphics[width=0.9\columnwidth]{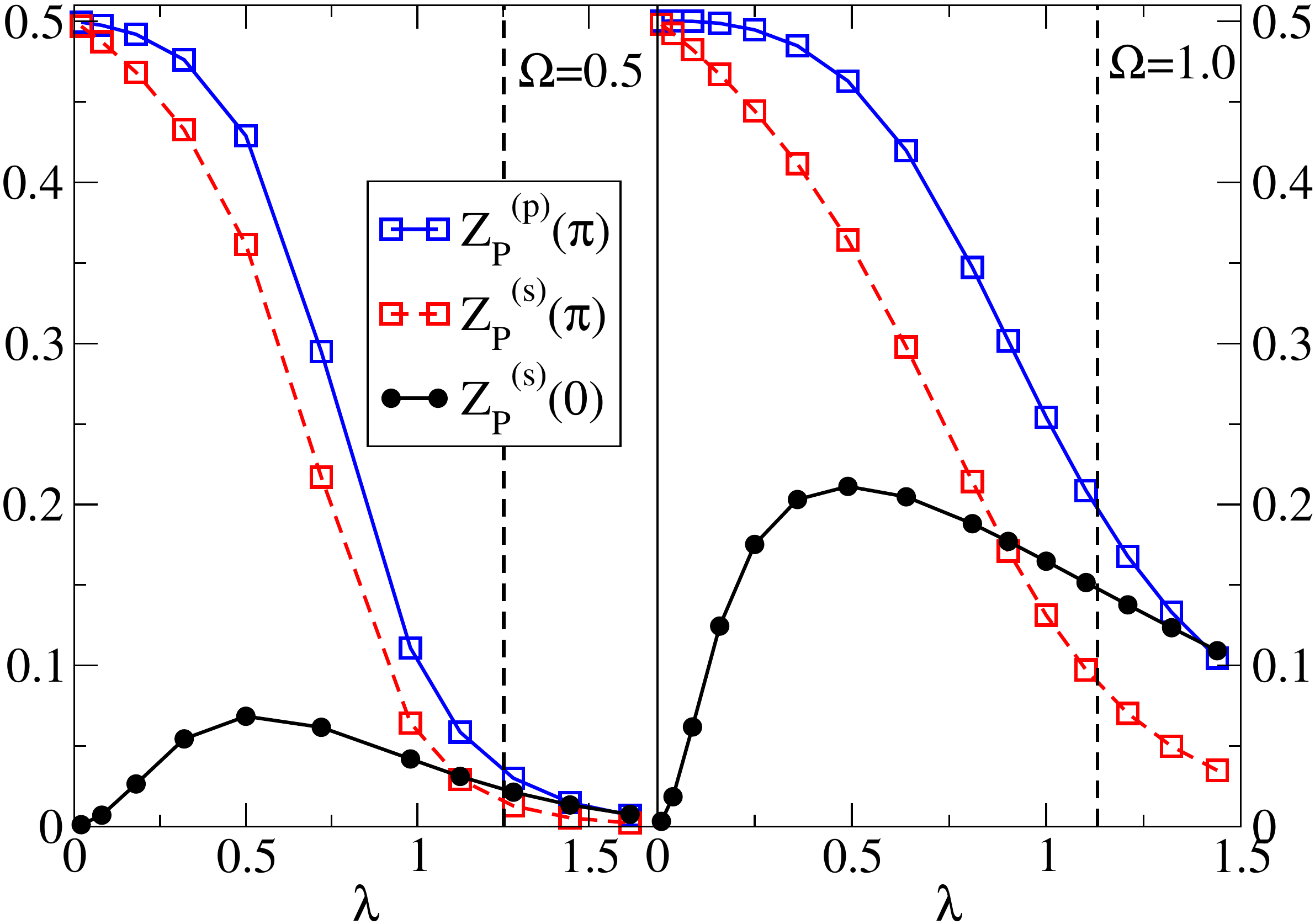} 
  \caption{(color online) The QP weight $Z_{\text{P}}^{(s)}$\ and
    $Z_{\text{P}}^{(p)}$ for $\Omega=0.5$ (left panel) and
    $\Omega=1.0$ (right panel). Note that $Z_{\text{P}}^{(p)}(0)=0$,
    because the $s$ and $p$-orbitals do not hybridize at $k=0$. The
    vertical, dashed line marks $\lambda_c$. Convergence was reached
    for $M=14,18$, respectively.}
  \label{fig:ZGS_vs_g}
\end{figure}

Changing $\Delta$ results in a shift of the polaron band. The polaron
energy must be smaller than that of the free hole, and the latter is
shifted downwards (upwards) for $\Delta>0$ ($\Delta<0$). A similar
shift is therefore expected at least for small $\lambda$, and it is
seen to appear for all $\lambda$. The effect of $\Delta$ on
$\lambda_c$ is discussed below.

Before discussing the nature of the transition in detail, we quickly
analyze the QP weight, $Z_{\text{P}}^{(s,p)}(k)$, defined as the
overlap between the polaron eigenstate and the $s$ and $p$ free-hole
states, respectively. They are shown in Fig. \ref{fig:ZGS_vs_g}. Note
that $Z_{\text{P}}^{(p)}(0) = 0$ for any finite $\lambda$, because the
$s$ and $p$ orbitals do not hybridize at $k=0$ (see Eq.
(\ref{eq:Heff_matrix})) and is therefore not shown. This is in stark
contrast to the free hole case, where at $k=0$ all the free-hole
weight is on the $p$-orbital. Consequently there is a discontinuous
change in the $k=0$  QP weight when the hole-phonon coupling is turned
on. 

For $k=\pi$ the QP weight falls off rapidly as $\lambda$ increases,
whereas for $k=0$ its projection on the $s$-orbitals first increases
and then falls off more slowly. As pointed out above, this initial
increase in QP weight occurs because at $k=0$ and for sufficiently
small $\lambda$, the polaron band lies just below the polaron+phonon
continuum.

The vertical dashed lines mark $\lambda_c$, the value of $\lambda$ at
which the $k=\pi$ and $k=0$ polaron energies cross. Note that there is no sudden change in the QP weights at
$\lambda_c$. This is a strong hint that the crossover is not due to a
change in the nature of the phonon cloud.

\begin{figure}[b]
  \centering \includegraphics[width=\columnwidth]{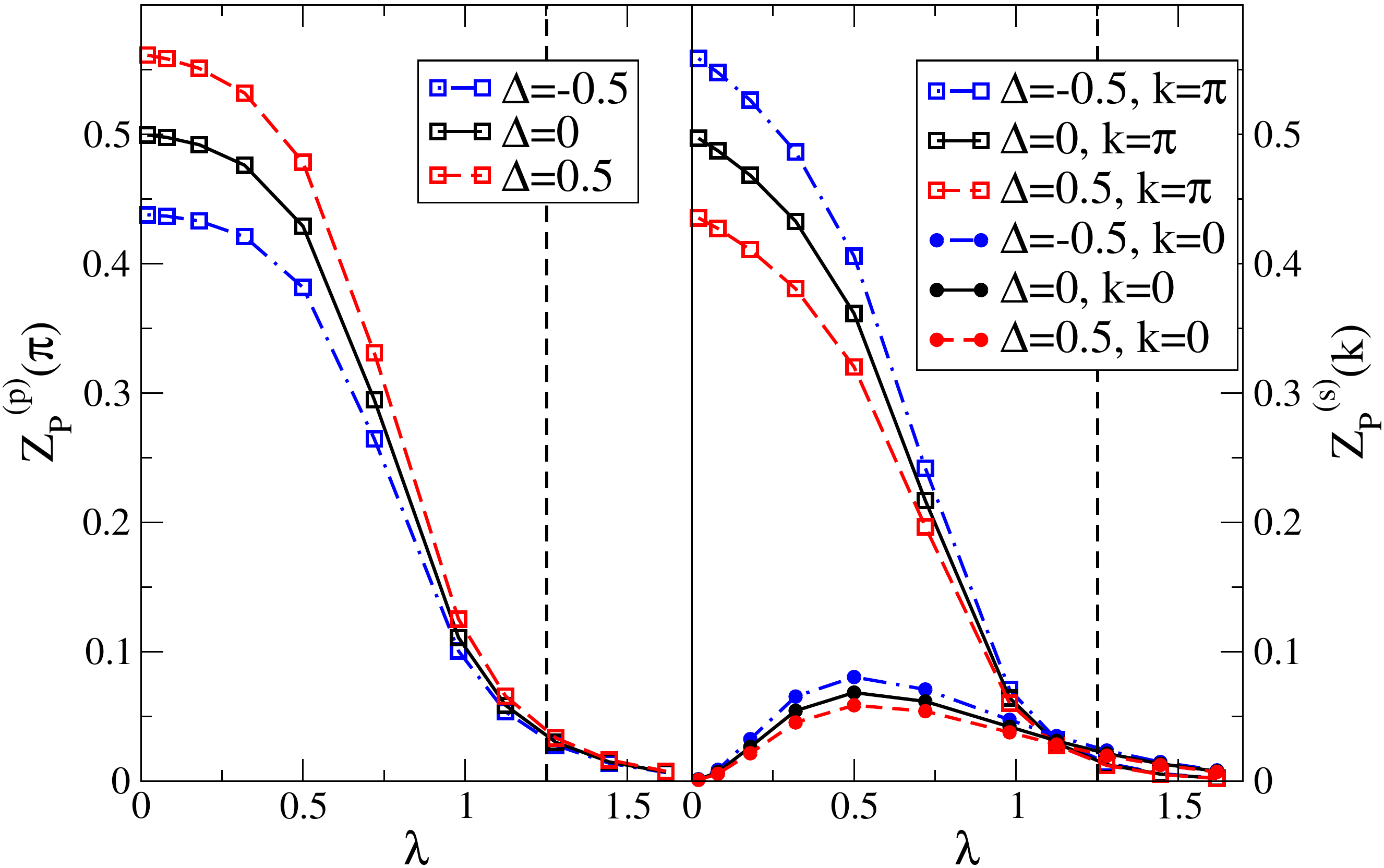}
  \caption{(color online) The QP weight $Z_{\text{P}}^{(p)}$ (left
    panel) and $Z_{\text{P}}^{(s)}$ (right panel) for $\Omega=0.5$.
    Note that $Z_{\text{P}}^{(p)}(0)=0$, because the $s$ and $p$-orbitals
    do not hybridize at $k=0$. The vertical, dashed line marks
    $\lambda_c$. Convergence was reached for $M=18$}
  \label{fig:ZGS_vs_g_diff_delta}
\end{figure}

The dependence of the QP weight on $\Delta$ is shown in Fig.
\ref{fig:ZGS_vs_g_diff_delta} for $\Omega=0.5$. As expected,
$\Delta>0$ increases the amount of $p$ character and decreases the $s$
character; $\Delta<0 $ has the opposite effect. For larger
values of the coupling $\lambda$, the change in QP
weights due to $\Delta$ becomes negligibly small. At these values of
$\lambda$ the phonon cloud is already quite sizable, as indicated by
the small values of the QP weights. Together, Fig.
\ref{fig:EGS_vs_g} and Fig. \ref{fig:ZGS_vs_g_diff_delta} show that
for strong coupling, $\Delta$ shifts the energy of the
polaron but does not change the nature of its phonon cloud. This, in turn, suggests
that $\hat{H}_{\text{h-ph}}$ strongly favors a specific
kind of phonon cloud. In other words, there appears to be only one dominant
mechanism that allows the hole to lower its energy via the emission
and absorption of phonons.

The (shifted) polaron dispersion $E_{\text{P}}(k)-E_{\text{P}}(0)$ vs $k$, close to the critical coupling $g_c$, is shown
in Fig. \ref{fig:dispersion_close_to_gc} for $\Omega=0.5, 1.0$ and
$100$.  Note that close to $g_c$, the polaron bandwidth is extremely
narrow. These results show that the GS momentum does indeed change discontinuously
from $\pi$ to $0$, justifying our definition of $g_c$.

For $\Omega=100$, the prediction of Eq. (\ref{eq:Eeff}) (dashed lines)
agrees quite well with ED and clearly reproduces the transition. As we
show now, a closer examination of this perturbative result indeed
unveils a lot of the physics governing this transition. When $\lambda
\ll t$, the polaron dispersion given by Eq. (\ref{eq:Eeff}) is not
much different from the free hole dispersion
$E_{\text{P}}^{\text{eff}}(k) \approx E_0(k)$, with the ground state
at $k=\pi$. For $\lambda \gg t$, on the other hand, the polaron
dispersion can be approximated as:
\begin{align}
  E_{\text{P}}^{\text{eff}}(k) \approx -t \lambda \left [2 + \cos(ka)
    + \left |\cos(ka) - \frac{\Delta}{2t\lambda} \right | \right] -
  \frac{\Delta}{2},
\end{align}
with the ground state at $k=0$ and a flat band for $ka \geq
\arccos(\Delta/(2t\lambda))$. Consequently there is a critical value,
$\lambda_c$, where the ground state momentum changes. Indeed it can be
verified that for
\begin{align}
  \lambda_c = \frac{\Delta}{4t} + \sqrt{\frac{\Delta^2}{16 t^2} +
    \frac{1}{2}},
  \label{eq:lambda_c}
\end{align}
the $k$-dependence in the square root of Eq. (\ref{eq:Eeff}) exactly
cancels the $-t \lambda \cos(ka)$ term and the band becomes completely
flat, $E_{\text{P}}^{\text{eff}}(k) = -4t\lambda_c$. For $\lambda <
\lambda_c$ the ground state is at $k=\pi$, and for $\lambda >
\lambda_c$ it is at $k=0$.

It is clear that this change of the ground state momentum is due to a
competition between the antisymmetric, bare hopping $\hat{T}$ and the
symmetric, phonon-modulated hopping, $\hat{H}_{\text{h-ph}}$. To our
knowledge this is the first report of a change in the ground state
momentum caused by a symmetric phonon-modulated hopping;  such changes
have been  seen before only for models with antisymmetric,
phonon-modulated hopping.\cite{Zhao, DominicPRL} However, while for
the latter case the ground-state momentum changes {\em smoothly} for
$\lambda > \lambda_c$, here we observe a {\em discontinuous jump} from the
edge to the centre of the BZ. Note, furthermore, that the
hybridization between the two bands which leads to the square root in
Eq. (\ref{eq:Eeff}) plays an important role for the nature of the
transition and appears to be responsible for the flatness of the band
close to $\lambda_c$.

Although perturbation theory explains many  features of the
transition, there are some differences between the large $\Omega$ case
and
the cases where $\Omega$ is comparable to $t$. Eq. (\ref{eq:Eeff})
predicts a completely flat band at $\lambda_c$. While this is seen
for $\Omega=100$, it is no seen for smaller $\Omega$. Instead, here  higher order
processes leading to  phonon-mediated, longer range effective hopping, cause the
bandwidth to remain finite at all values of $\lambda$. We can
attribute these processes to longer range effective hopping because they lead to
a maximum at $k\approx\pi/2$, halving the BZ. Their contribution is
very small, as indicated by the  small bandwidth close to
$\lambda_c$. This leads us to conclude that the phonon-assisted
hopping process described in Section \ref{sec:perturbation-theory} is
indeed primarily responsible for most of the mobility of the polaron.

To make these arguments more compelling and valid for phonon
clouds with more than one phonon, we now analyze the nature of the
polaron cloud in more detail. For $k=0$,
the polaron has QP weight on $s_k^\dagger|0\rangle$ but not
$p_k^\dagger|0\rangle$. Acting with $\hat{H}_{\text{h-ph}}$ on
$s_{k=0}^\dagger|0\rangle$ (and ignoring normalization factors) gives
$\sum_j p_j^\dagger b_j^\dagger |0\rangle$, which we therefore expect
to be a part of the polaron wavefunction. Acting on this again with
$\hat{H}_{\text{h-ph}}$ gives $\sum_j s_j^\dagger[b_j^{\dagger,2} +
  b_{j-1}^{\dagger,2}] | 0\rangle$. This pattern continues and we find that for $k=0$ the 
polaron eigenfunction contains states of the type $\sum_j p_j^\dagger
b_j^{\dagger,2n+1} |0\rangle$, {\em i.e.} with an odd number of phonons
and the hole on the same $p$-orbital, and $\sum_j
s_j^\dagger[b_j^{\dagger,2n} + b_{j-1}^{\dagger,2n}] | 0\rangle$,
{\em i.e.} a symmetric state with an even number of phonons and the
hole on the adjacent $s$-orbitals. While we only considered
states with a one-site phonon cloud  adjacent to the hole,
this structure of the phonon cloud is indeed verified by ED, which finds very small weight for other configurations.

For $k=\pi$, $\hat{H}_{\text{h-ph}} s_\pi^\dagger |0 \rangle=0$.
Consequently we need to start the construction of the eigenstate from
$p_\pi^\dagger |0\rangle$. Acting with $\hat{H}_{\text{h-ph}}$ gives
$\sum_j (-1)^j s_j^\dagger(b_j^\dagger - b_{j-1}^\dagger)
|0\rangle$. Acting once more with $\hat{H}_{\text{h-ph}}$ gives
$\sum_j (-1)^j p_j^\dagger b_j^{\dagger,2} |0\rangle$. This pattern
generalizes and mixes states of the form $\sum_j (-1)^j p_j^\dagger
b_j^{\dagger,2n} |0\rangle$, {\em i.e.} with an even number of phonons
and the hole on the same $p$-orbital, and $\sum_j (-1)^j
s_j^\dagger[b_j^{\dagger,2n+1} - b_{j-1}^{\dagger,2n+1}] |0\rangle$,
{\em i.e.} an antisymmetric state with an odd number of phonons and
the hole on the adjacent $s$-orbitals. This also is verified by ED.

Thus, the polaron cloud structure is very different at $k=0$ and
$\pi$, and this has important consequences. Consider the
configuration $\sum_j \e{ikR_j} s_{j}^\dagger b_{j}^{\dagger,2n}
|0\rangle$. This state can be moved by first acting $2n$ times with
$\hat{H}_{\text{h-ph}}$; among other states, this links to $\sum_{j}
\e{ikR_j} s_{j+1}^\dagger |0\rangle$. Acting another $2n$ times with
$\hat{H}_{\text{h-ph}}$ gives (among many other states) $\sum_j
\e{ikR_j} s_{j+1}^\dagger b_{j+1}^{\dagger,2n} |0\rangle$, {\em i.e}
the original state translated by one site. Clearly, such processes
contribute to the mobility of the polaron. Similarly, states like
$\sum_j \e{ikR_j} p_{j}^\dagger b_{j}^{\dagger,2n+1} |0\rangle$ can
be moved by first acting $2n+1$ times with $\hat{H}_{\text{h-ph}}$
linking to $\sum_j \e{ikR_j} s_{j+1}^\dagger |0\rangle$, and then to
$\sum_j \e{ikR_j} p_{j+1}^\dagger b_{j+1}^{\dagger,2n+1} |0\rangle$
after applying $\hat{H}_{\text{h-ph}}$ another $2n+1$ times. For these
processes to work, it is crucial that the number of phonons is even (odd) if
the carrier is on an $s (p)$ orbital. This is the case for $k=0$, but for $k=\pi$ we
found that exactly the opposite is the case. This explains why the
phonon-modulated effective hopping, which dominates at large
couplings, favours $k=0$ and this eventually becomes the
ground-state. Note, furthermore, that this discussion also illustrates why
an MA version restricted to a one-site phonon cloud is expected to
capture well the phenomenology of this model.

\begin{figure}[t]
  \centering \includegraphics[width=\columnwidth]{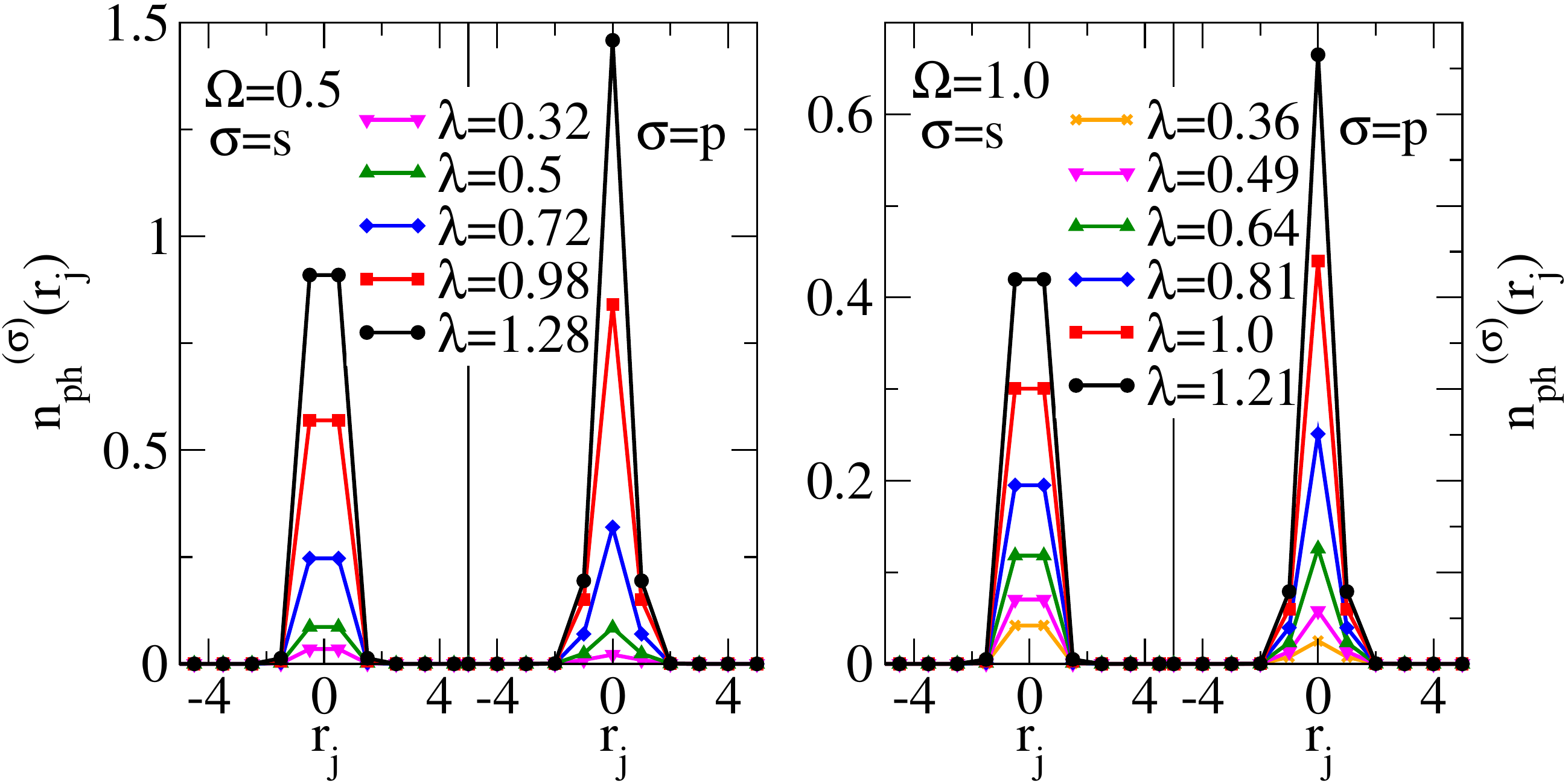}
  \caption{(color online) The phonon distribution
    $n_{\text{ph}}^{(\sigma)}(r_j)$ defined in Eq. (\ref{eq:nph}) for
    $\Omega=0.5$ (left panel) and $\Omega=1.0$ (right panel) at
    different values of $\lambda$. Other parameters are $k=\pi$ and
    $\Delta=0$. Convergence was reached for $M=14$ and $18$,
    respectively.}
  \label{fig:nph}
\end{figure}

To analyze the spatial extent of the polaron cloud we calculate the
phonon distribution
\begin{align}
  n_{\text{ph}}^{(\sigma)}(r_j) = \sum_i \langle \text{P}_\pi |
  c_{i,\sigma}^\dagger c_{i,\sigma} b_{i+j}^\dagger b_{i+j}
  |\text{P}_\pi \rangle,
  \label{eq:nph}
\end{align}
where $| \text{P}_\pi \rangle$ is the polaron-state with momentum
$k=\pi$ and $r_j$ is the distance between the carrier at site $i$ and
the phonon at site $i+j$. This means that $r_j = j$ when the carrier
is on a $p$-orbital, and $r_j = j+1/2$ when it is on an
$s$-orbital. The results are shown in Fig. \ref{fig:nph} for
$\Omega=0.5$ and $\Omega=1.0$. We see that the polaron cloud
is located in the immediate vicinity of the carrier, {\em i.e.} we are
dealing with a very small polaron. If the carrier is on an $s$-orbital
the majority of phonons are hosted by the two oxygens next to it,
while for a carrier on a $p$-orbital the three closest oxygen sites
contribute. As $\lambda$ is increased past the critical coupling
($\lambda_c = 1.25$ for $\Omega=0.5$ and $\lambda_c=1.13$ for
$\Omega=1.0$), there is no noticeable change in the nature of the
cloud. We also find that for $k=0$ and sufficiently large
$\lambda$ the phonon cloud is qualitatively similar (not shown) to that for
$k=\pi$. For small $\lambda$ this is not true
as here the polaron state at $k=0$ is not well-separated from the
polaron + phonon continuum. Reducing $\Omega$ 
drastically changes the overall number of phonons, but not the spatial
extent of the cloud.

\begin{figure}
  \centering \includegraphics[width=\columnwidth]{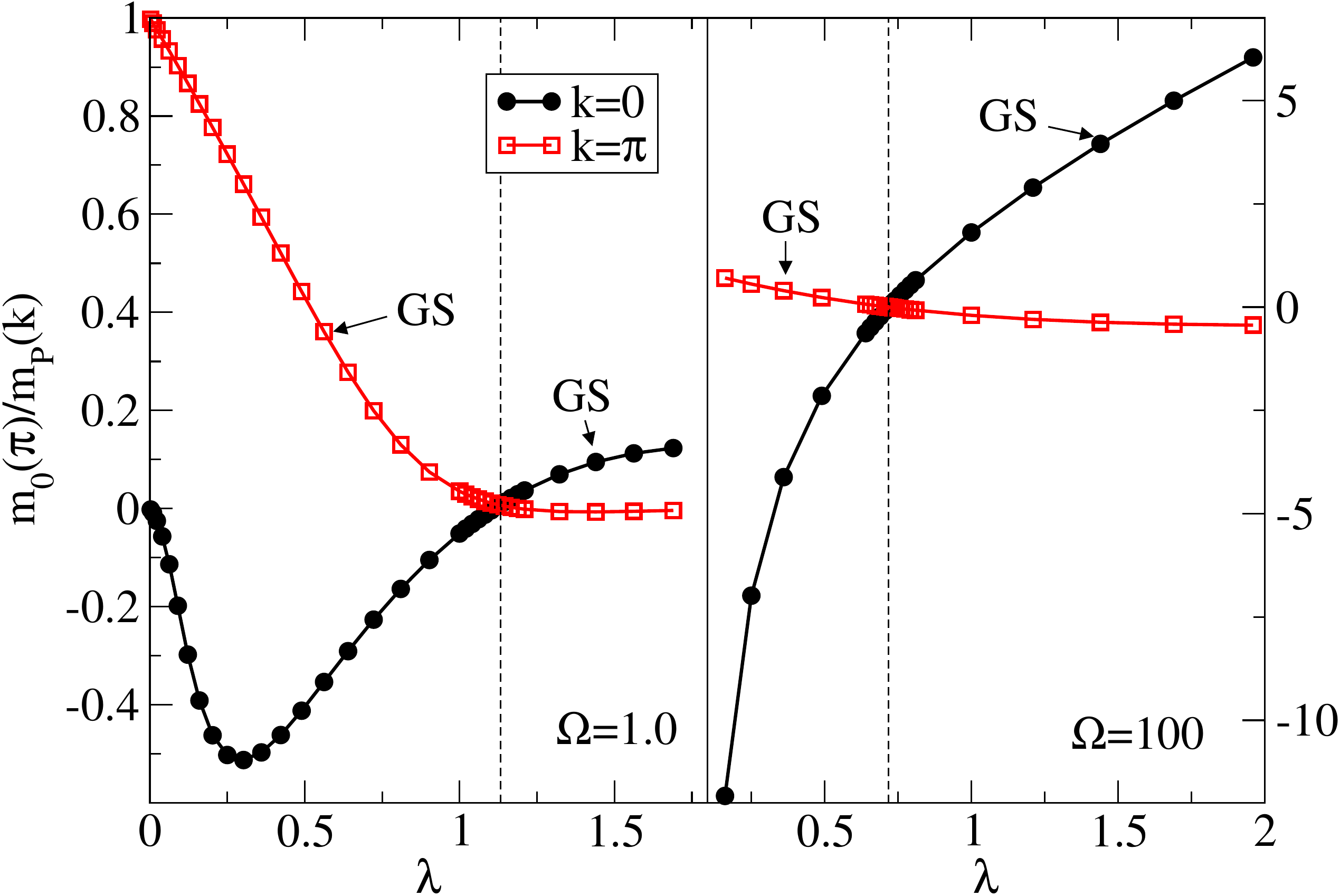}
  \caption{(color online) Inverse effective polaron mass $m_{\text{P}}^{-1}(k)$ normalized 
    by the free hole mass $m_{0}^{-1}(\pi)$ for $\Omega=1.0$ (left panel)
    and $\Omega=100$ (right panel) and $\Delta=0$. The dashed line marks
    $\lambda_c$. The values of $M$ are 14 and 10, respectively.}
  \label{fig:meff}
\end{figure}

Apart from the small bandwidth, the heavy nature of the polaron also
becomes apparent from the value of its inverse effective mass,
$m_{\text{P}}^{-1}(k)=\partial^2 E_{\text{P}} /\partial k^2$, shown in
Fig. \ref{fig:meff} for $k=0,\pi$ at $\Omega=1.0$ and
$\Omega=100$. Note that we normalize this with the free hole mass.  As
$\lambda\rightarrow\lambda_c$ from below, $m_{\text{P}}^{-1}(\pi)$
decreases indicating that the polaron becomes heavier. This is
expected because as the discussion above shows, phonon-modulated
hopping cannot move the polaron cloud at this momentum, and the bare
hole hopping is renormalized to smaller values due to the presence of
the phonon cloud. At $\lambda=\lambda_c$ the ground state changes
momentum and consequently we now need to follow $m_{\text{P}}^{-1}(0)$
which increases because here the phonon-modulated hopping is active
and increases the mobility of the polaron.

For large values of $\Omega$ the increase of $m_{\text{P}}^{-1}(0)$
for $\lambda>\lambda_c$ is substantial resulting in a much lighter
polaron than at $\lambda<\lambda_c$. For smaller values of $\Omega$,
on the other hand, $m_{\text{P}}^{-1}(0)$ levels off quite rapidly and
the polaron remains heavy. This is to be expected because for smaller
$\Omega$, the phonon cloud is quite large at $\lambda_c$, suppressing
the polaron's mobility. For $\Omega \gg t$, on the other hand, the
average phonon number is small and an increase in $\lambda$ results in
an increase of the effective hopping between $s$-orbitals. This
effective hopping can become larger than the direct hopping, resulting
in a lighter polaron.
 
A major difference between the two panels of Fig.  \ref{fig:meff} is
that for $\Omega=1.0$ the inverse effective mass at $k=0$ has a
minimum at finite $\lambda$. Again, this is due to the presence of the
polaron+phonon continuum, which for sufficiently small $\lambda$
forces the polaron band to flatten out near $k=0$. Indeed, the location of
this minimum agrees with the location of the maximum in
$Z_{\text{P}}^{(s)}(0)$ in Fig. \ref{fig:ZGS_vs_g}.

\begin{figure}
  \centering \includegraphics[width=\columnwidth]{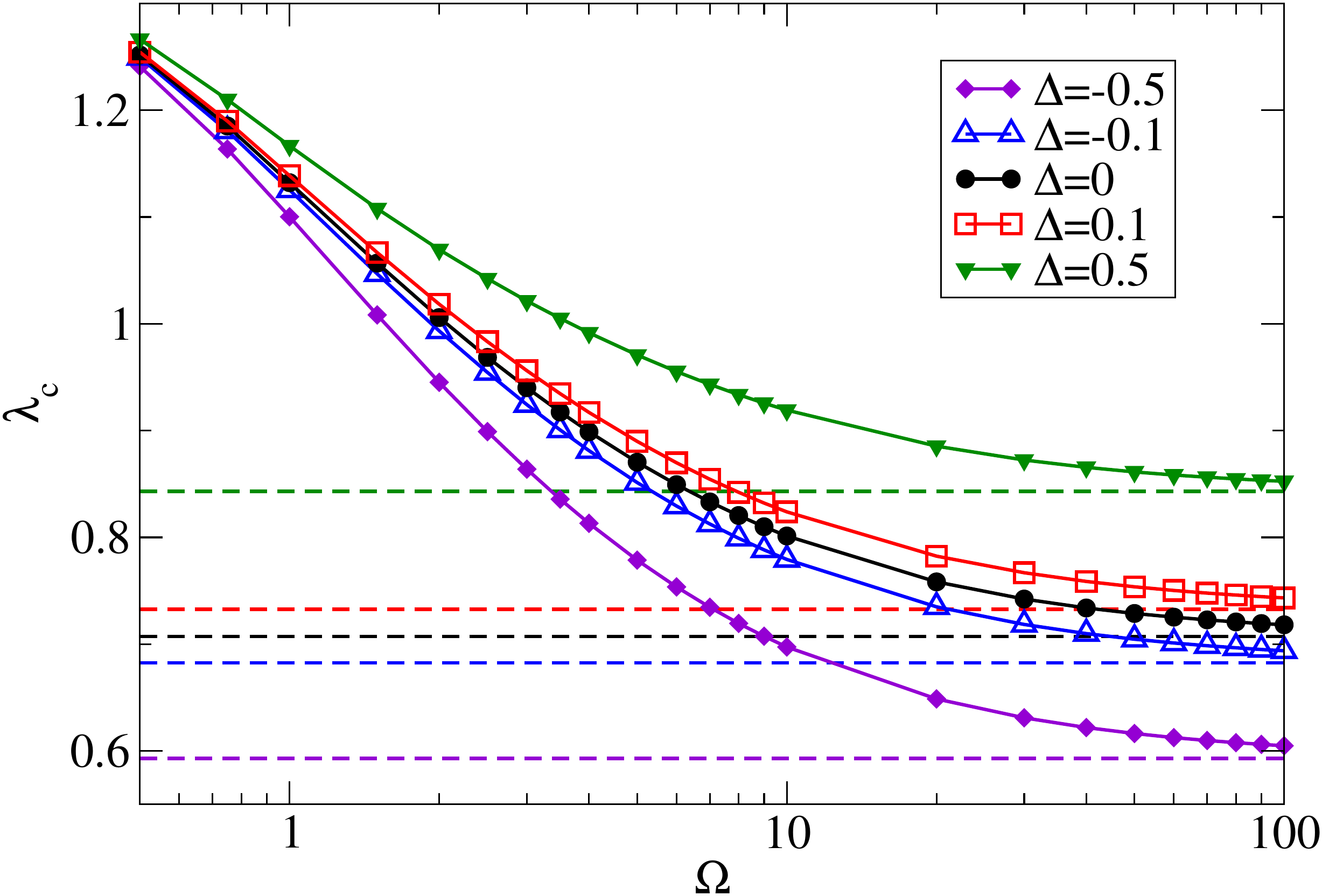} 
  \caption{(color online) $\lambda_c$ as a function of $\Omega$ and $\Delta$.
    Different values of $M$ were used to ensure convergence. Dashed
    lines are the perturbative result from Eq. (\ref{eq:lambda_c}).}
  \label{fig:lambda_c}
\end{figure}

Let us now discuss the role of the charge transfer energy $\Delta$.
In Fig. \ref{fig:lambda_c} we show the critical coupling $\lambda_c$
for different values of $\Delta$ and $\Omega$. The dashed lines show
the perturbative result, Eq. (\ref{eq:lambda_c}), which for
sufficiently large $\Omega$ is in good agreement with the ED results.

\begin{figure}
  \centering \includegraphics[width=\columnwidth]{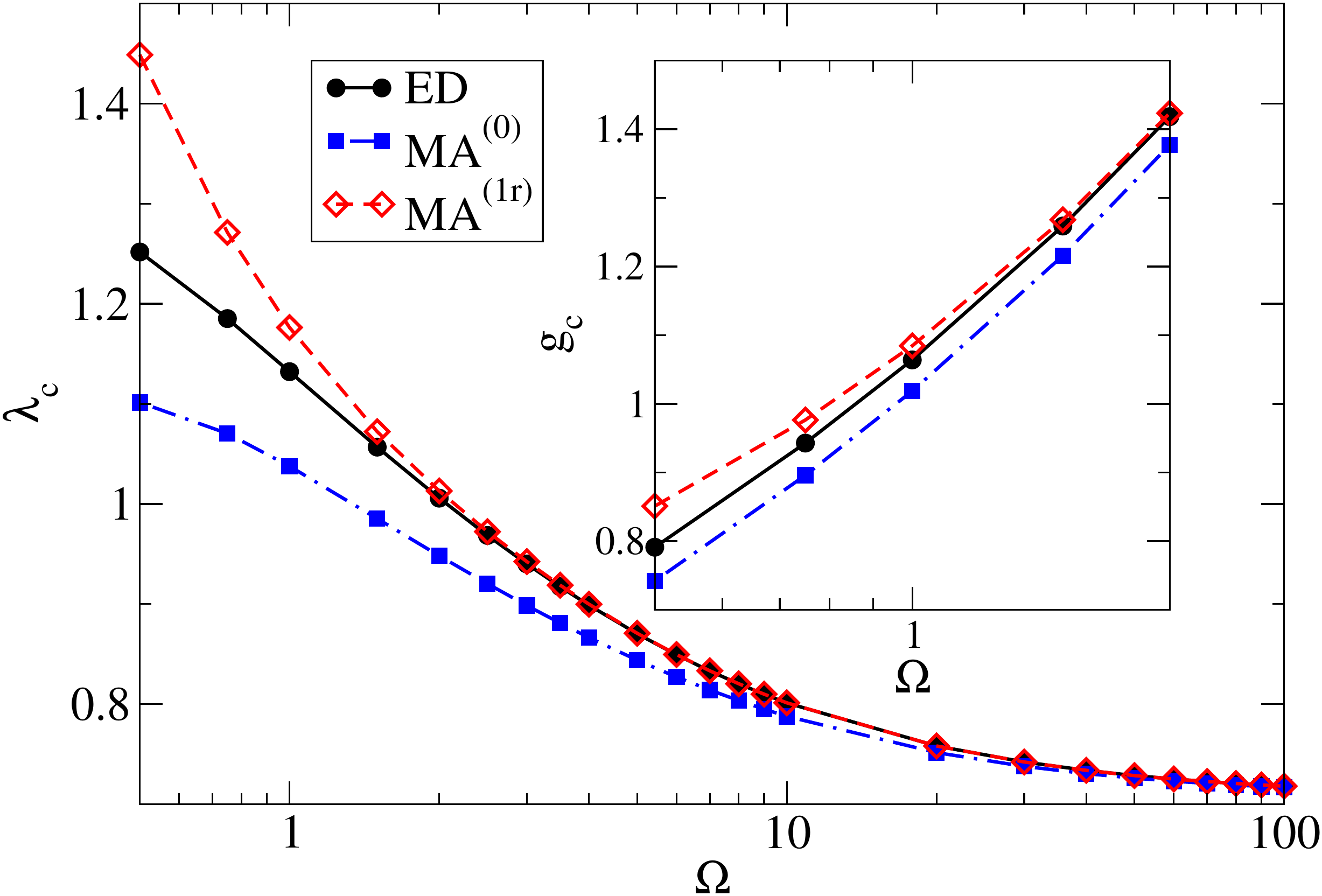}
  \caption{(color online) Comparison between the exact value of
    $\lambda_c$ obtained with ED for $\Delta=0$, and the predictions
    of MA$^{(0)}$ and MA$^{(1r)}$.}
  \label{fig:MA_gc}
\end{figure}

The critical coupling increases with increasing $\Delta$. This is not
surprising since a large value of $\Delta$ favors the $p$ character,
whereas the phonon-modulated hopping favors the $s$ character.
Consequently, a negative value of $\Delta$ facilitates the transition.
However, this effect is significant only for relatively large values
of $\Omega$. As $\Omega$ decreases the variation of $\lambda_c$ with
$\Delta$ becomes very small, suggesting that here $\Delta$ has little
influence on the nature of the polaron cloud. This agrees with the
conclusions drawn from Fig. \ref{fig:ZGS_vs_g_diff_delta}. At small
values of $\Omega$ the value of $\lambda_c$\ is also much larger than
that predicted by perturbation theory. However, $\lambda \propto
1/\Omega$ and in fact the transition from $k=0$\ to $k=\pi$ is
actually achieved at smaller values of $g$ for small $\Omega$ (see inset of Fig. \ref{fig:MA_gc}). For
these values of $\Omega$ the phonon cloud is quite large and therefore
it is to be expected that the bare hopping, favoring $k=\pi$, is renormalized substantially and the phonon-modulated hopping, favoring $k=0$, wins already at smaller values of $g$. 

\begin{figure}
  \centering \includegraphics[width=\columnwidth]{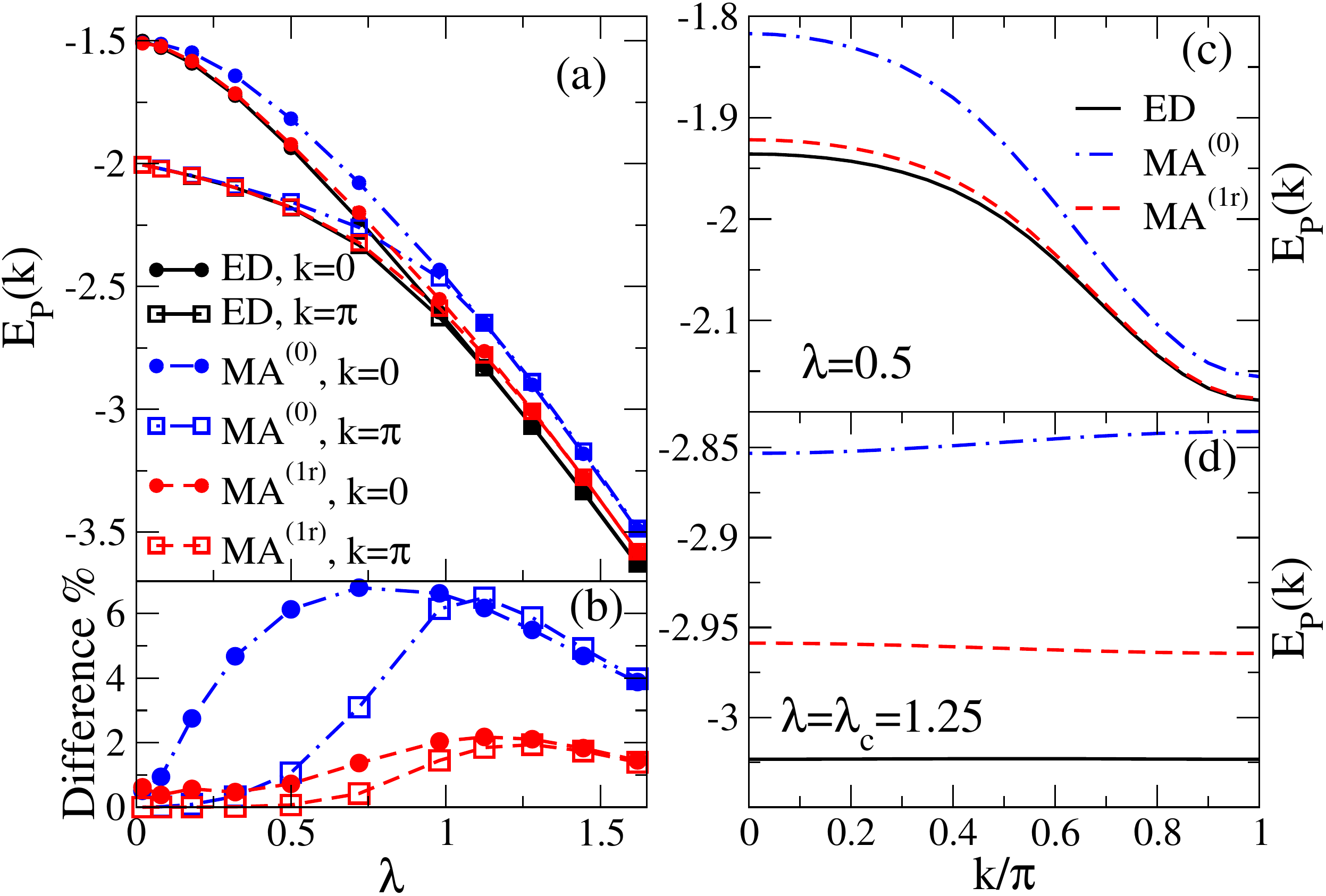}
  \caption{(color online) Comparison between ED, MA$^{(0)}$ and
    MA$^{(1r)}$. (a) $E_{\text{P}}(k)$ for $k=0, \ \pi$ at different
    values of $\lambda$. (b) The absolute value of the difference in
    \%. (c) The dispersion $E_{\text{P}}(k)$ at $\lambda=0.5$. (d)
    $E_{\text{P}}(k)$ at $\lambda=\lambda_c=1.25$. In all cases
    $\Omega=0.5$ and $\Delta=0$.}
  \label{fig:MA_1}
\end{figure}

This concludes our analysis of the ED results. We now briefly compare
the MA predictions with the ED results, in order to validate our
choice of the variational space. This is useful because MA can be much
more easily and efficiently extended to higher dimensions than ED
calculations. Moreover, prior work\cite{Mona-MA-1, Mona-MA-2} has
shown that any version of MA becomes more accurate in higher
dimensions, where the bare propagators decay faster with distance
outside the free-hole continuum. Note also that MA accuracy improves with increasing phonon
frequency $\Omega$;\cite{Mona-MA-1, Mona-MA-2} this is why we present results for $\Omega=0.5$
where the phonon cloud is very large at strong couplings, posing a
challenging test for this (and any other) approximation.

In panel (a) of Fig. \ref{fig:MA_1} we compare the MA results for the
polaron energies $E_{\text{P}}(\pi),E_{\text{P}}(0)$ at different values of
$\lambda$, to the ED results. As expected for a variational approach,
both MA$^{(0)}$ and MA$^{(1r)}$ values are always larger than the ED
ones.  Panel (b) of Fig. \ref{fig:MA_1} shows the
relative difference between MA and ED. At small values of
$\lambda$ both MA$^{(0)}$ and MA$^{(1r)}$ are very accurate, but at
larger values of $\lambda$, MA$^{(1r)}$ is clearly superior, showing
that its additional configurations acquire finite weight. Obviously,
adding more cloud configurations will further increase accuracy, but
it is clear that this rather small set already suffices to
capture quantitatively quite accurately the polaron properties.

A comparison for the dispersion $E_{\text{P}}(k)$ is shown in panel
(c) of Fig. \ref{fig:MA_1} for the intermediate coupling $\lambda=0.5$
and in panel (d) for the critical coupling $\lambda_c=1.25$. Panel (c)
shows that both MA$^{(0)}$ and MA$^{(1r)}$ give the best results for
$k=\pi$. This is probably due to the fact that at $\lambda=0.5$ the GS
is still at $k=\pi$, and a variational method like MA is expected to
perform best for the GS. Note that panel (b) also shows that for
$\lambda < \lambda_c$ the relative error is smaller for $k=\pi$ than
for $k=0$.

The dispersion shown in panel (d) is very narrow (on this scale). MA$^{(0)}$ and
MA$^{(1r)}$ both reproduce the small bandwidth quite well, but shifted to
higher energies by 2\% and 6\%, respectively. However, neither
MA$^{(0)}$ nor MA$^{(1r)}$ predict the exact value of
$\lambda_c=1.25$. Instead, MA$^{(0)}$ predicts a smaller value of 1.1,
whereas MA$^{(1r)}$ predicts a larger value of 1.45.  This trend is
true for all values of $\Omega$, as shown in Fig.
\ref{fig:MA_gc}. Here we also see that MA$^{(1r)}$
gives very good predictions for $\Omega > 1$. The agreement becomes
gradually worse as $\Omega \rightarrow 0$, where because of the low
cost of phonons, the spatial extent of the phonon cloud increases
beyond two sites. It also needs to be pointed out that for small
values of $\Omega$ small differences in the hole-phonon coupling $g$
are amplified in the effective coupling $\lambda = g^2/\Omega$. For
small $\Omega$ we therefore show the values of $g_c$ in the inset of
Fig. \ref{fig:MA_gc}.

\begin{figure}[b]
  \centering \includegraphics[width=\columnwidth]{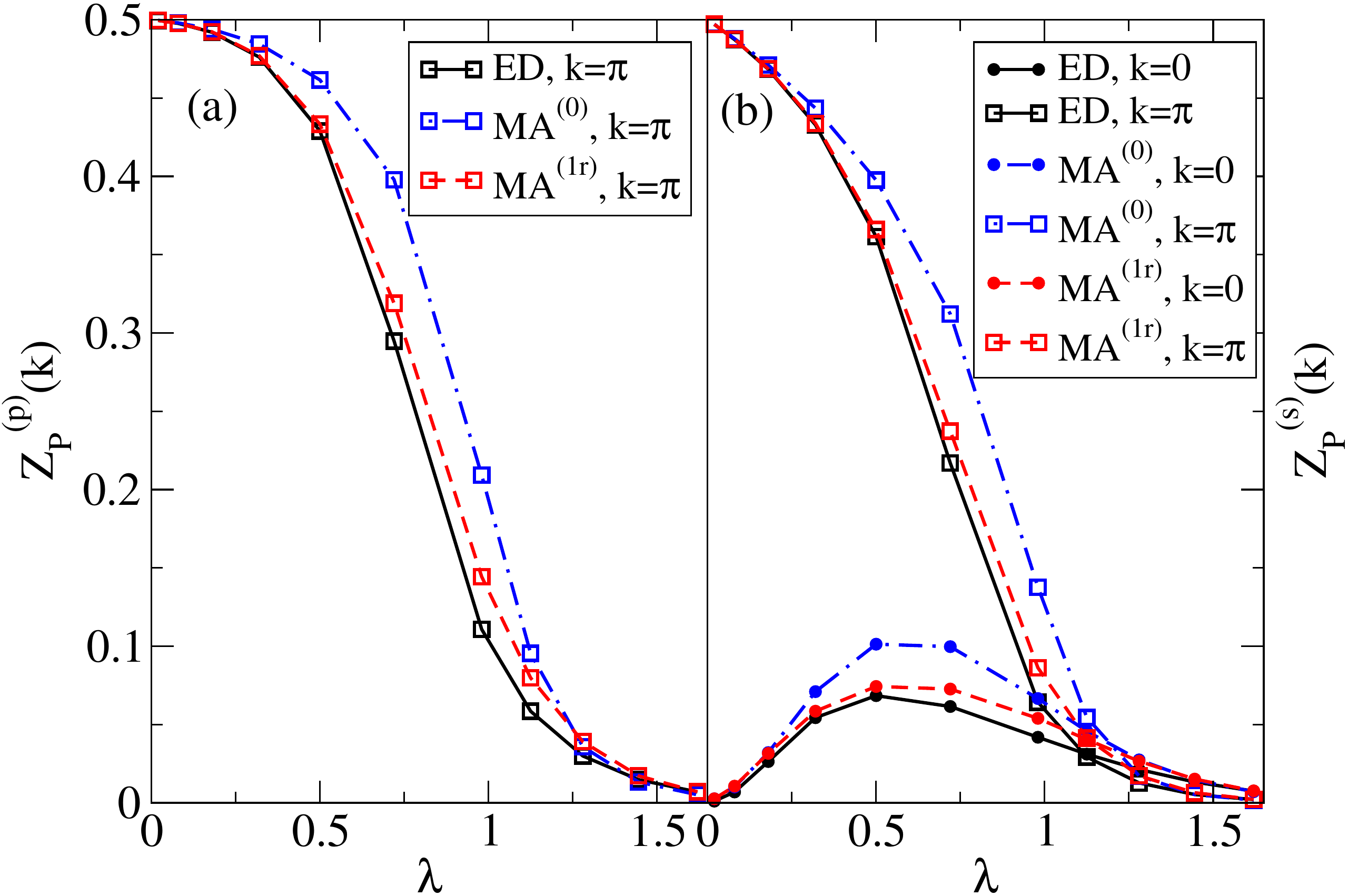}
  \caption{(color online) Comparison between ED, MA$^{(0)}$ and
    MA$^{(1r)}$. (a) The QP weight $Z_{\text{P}}^{(p)}(k)$ for $\pi$
    for different values of $\lambda$. Note that
    $Z_{\text{P}}^{(p)}(0)=0$ for all $\lambda$. (b)
    $Z_{\text{P}}^{(p)}(k)$ for $k=0,\ \pi$. In all cases $\Omega=0.5$
    and $\Delta=0$}
  \label{fig:MA_2}
\end{figure}

A comparison of the QP weights obtained with MA and ED is shown in
Fig. \ref{fig:MA_2}. Panel (a) shows $Z_{\text{P}}^{(p)}(0)$ (as
pointed out above, $Z_{\text{P}}^{(p)}(\pi)=0$ for all
$\lambda$). Panel (b) shows $Z_{\text{P}}^{(s)}(k)$ for $k=0$ and
$k=\pi$.  As expected for a variational approximation, the MA weights
are larger than the ED values everywhere.

For small $\lambda$ both MA$^{(0)}$ and MA$^{(1r)}$ perform very well.
The largest quantitative disagreement appears for intermediate values
of $\lambda$. Here, $Z_{\text{P}}^{(s,p)}(\pi)$ falls off quite
rapidly, and this is captured only qualitatively by MA. Note that
MA$^{(1r)}$ performs much better than MA$^{(0)}$ in this
$\lambda$-range, indicating that a further improvement of MA by adding
phonons on adjacent sites is possible. The results improve
substantially for $\lambda > \lambda_c$. At these values, the QP
weight does not change as rapidly anymore. We find that here
MA$^{(0)}$ and MA$^{(1r)}$ are on equal footing, with MA$^{(0)}$
actually outperforming MA$^{(1r)}$ at $\lambda=1.45$. This could be
related to the fact that MA$^{(1r)}$ predicts a value that is too
large for $\lambda_c$, whereas MA$^{(0)}$ predicts a value that is
smaller than the exact result.

\section{Conclusions}
\label{sec:conclusions}

We have investigated the effect of phonon-modulated hopping in a
two-band model describing a chain with alternating atoms with $s$ and
$p$ valence orbitals, respectively. As discussed in the Introduction,
part of the motivation for this work is to study a toy 1D model
inspired by the perovskite LaBiO$_3$ (or nearly fully compensated
BaBiO$_3$) in the scenario where holes are primarily located on the O
atoms, to understand the properties of the resulting polaron and to
develop a good approximation that can be easily generalized for
similar 3D system.

The key finding is that at sufficiently large hole-phonon coupling,
the polaron undergoes a sharp transition where its GS momentum jumps
from $\pi$ to $0$. Sharp polaron transitions have been observed
previously in the SSH and related models with phonon-modulated
hopping, \cite{DominicPRL,Zhao,MonaPRL110} however there are
qualitative differences between our results and those of these
studies. In these other models the momentum changes {\em smoothly} as
the effective coupling is increased. Also, the polaron mass diverges
at the transition but remains finite (and can be surprisingly light)
in the limit of infinite effective coupling. In our model, on the
other hand, there is a {\em discontinuous} jump of the momentum from
$k=\pi$ to $k=0$. To the best of our knowledge, such a discontinuous
transition has not been discussed before. These results suggest that
there may be more new physics to be discovered in models of
carrier-phonon coupling different from that of the much studied
one-band models of the Holstein and Fr\" ohlich type.

We argued that this transition is due to a competition between the
bare hopping $\hat{T}$ and the phonon-modulated hopping
$\hat{H}_{\text{h-ph}}$: the former favors a $k=\pi$ ground state,
while the latter increases polaron mobility at $k=0$ but not at
$k=\pi$. This analysis suggests that a transition can only occur when
$\hat{T}$ and $\hat{H}_{\text{h-ph}}$ have different symmetries. This
is supported by the findings of Zhang {\em et. al.}  \cite{Zhao} who
report a transition in an SSH-like model for symmetric bare hopping
and antisymmetric phonon-modulated hopping, but not for symmetric
phonon-modulated hopping.

We have used arguments based on perturbation theory, but also more
general symmetry arguments, to support the view that the discontinuous
nature of this transition is closely tied to the two-band nature of
the model. In perturbation theory the hybridization between the $s$
and $p$ bands leads to the square root dependence in Eq.
(\ref{eq:Eeff}), which is the main ingredient needed for this type of
discontinuous jump of the GS momentum. We believe that it is highly
unlikely that a transition with such a discontinuous jump can be found
in a one-band model. This suggests that the polaron properties of
multi-band models are qualitatively different from those of one-band
models, and that more work needs to be done to exhaustively study all
possible types of polaron transitions. Whether such transitions also
occurs in 2D and 3D models where nn hopping between oxygens is also
included is not a priori clear; we believe this to be the case and we
are currently investigating such models.

When extended to one-site phonon cloud configurations with an
arbitrary number of phonons, the arguments mentioned above allowed us
to elucidate the mechanism behind the transition. They also suggested
two types of MA variational approximations, where the phonon cloud is
restricted to be one-site in extent (MA$^{(0)}$) and one-site cloud
with at most one additional phonon on an adjacent site
(MA$^{(1r)}$). We find that MA$^{(0)}$ reproduces the GS energy with
an accuracy of 6\% or better while MA$^{(1r)}$ yields even better
results with an accuracy of at least 2\%. These values were found for
$\Omega=0.5$ where the phonon cloud is already very large, and
therefore show that MA performs well even under difficult conditions.

Of course, the MA accuracy is significantly less for the QP weights at
intermediate coupling $\lambda$. This is not surprising, because the
part of the eigenstates ignored by the variational calculation does
contribute to the wavefunctions' normalization (and therefore serves
to lower the QP weight) even if it may not have much effect on the
eigenenergy. It is important to note, however, that both MA approaches
reproduce the qualitative behavior correctly, and that adding more
variational states will further improve quantitative agreement. MA is
also successful in predicting an accurate value for the critical
coupling $\lambda_c$ for not too small phonon frequencies. At small
$\Omega$, $\lambda_c$ is underestimated by MA$^{(0)}$ and
overestimated by MA$^{(1r)}$, so at least they provide bounds for its
true value.

The major advantage of MA, when compared to ED, is its numerical cost.
For MA$^{(0)}$ one merely needs to multiply $2 \times 2$-matrices,
whereas for MA$^{(1r)}$ the largest matrix size is $17 \times 17$. In
fact one can also store all of the $\alpha_n$ and $\beta_n$ in one
large sparse matrix, which further increases performance. The
essential aspect is that in both cases the number of states for a
given number $n$ of phonons is independent of $n$. This is why
generalizing these MA approaches to 3D will be significantly more
efficient than for more general MA schemes, like that of Ref.
\onlinecite{Glen-BM}. We have shown  that in 1D these schemes
already work quite well, and since MA  improves its
accuracy in higher dimension, these generalizations should suffice to accurately and efficiently obtain polaron properties in the multi-band, 3D model. This work is now in progress.

\begin{acknowledgements}
We thank George Sawatzky for suggesting this problem, and C.P.J.
Adolphs for sharing his experience with ED. This work was supported by
NSERC, QMI and the UBC 4YF (M.M.M.).
\end{acknowledgements}

\appendix
\section{MA equations of motion}
The equations of motion (eom) for $G^{\sigma'\sigma}(k,\omega)$ can be
constructed by repeatedly using Dyson's identity $\hat{G}(\omega) =
\hat{G}_0(\omega)+\hat{G}(\omega) \hat{V} \hat{G}_0(\omega)$, where
$\hat{H}=\hat{H}_0+\hat{V}$ and $ \hat{G}_0(\omega)$ is the resolvent
for $\hat{H}_0$. Choosing $\hat{V}=\hat{H}_{\text{h-ph}}$ and doing
this once yields:
\begin{align}
  G^{\sigma'\sigma}(k,\omega) = &G_0^{\sigma'\sigma}(k,\omega) + g \left
  ( 1+\e{ika} \right )f_{p,0}^{(1)}(k,\omega) G_0^{s\sigma}(k,\omega)
  \nonumber \\ &+ \left ( f_{s,0}^{(1)}(k,\omega) + f_{s,1}^{(1)}
  \right) G_0^{p\sigma}(k,\omega),
\label{eq:eom_G}
\end{align}
where we defined the  generalized GFs:
\begin{align}
  f_{\sigma, l}^{(n)}(k,\omega) = \sum_j \frac{\e{ik R_j}}{\sqrt{N}}
  \langle 0 | c_{k,\sigma'} \hat{G}(\omega) c_{j+l,\sigma}^\dagger
  (b_{j}^\dagger)^n | 0 \rangle.
\label{eq:def_f}
\end{align}
These  describe the projection onto states with a
one-site phonon cloud, and are therefore the only GFs included in
MA$^{(0)}$. Note that we suppressed the $\sigma'$ index in the
definition of the generalized GFs since the eom do not couple to GFs
with different $\sigma'$. For MA$^{(1r)}$, we also include the
contribution from the generalized GFs:
\begin{align}
\nonumber  &F_{\sigma, l}^{(1,n)}(k,\omega) = \sum_j \frac{\e{ik R_j}}{\sqrt{N}}
  \langle 0 | c_{k,\sigma'} \hat{G}(\omega) c_{j+l,\sigma}^\dagger
  b_{j-1}^\dagger(b_{j}^\dagger)^n | 0 \rangle\\
\nonumber &F_{\sigma, l}^{(n,1)}(k,\omega) = \sum_j 
  \frac{\e{ik R_{j+1}}}{\sqrt{N}}
  \langle 0 | c_{k,\sigma'} \hat{G}(\omega) c_{j+l,\sigma}^\dagger
  (b_{j}^\dagger)^nb_{j+1}^\dagger | 0 \rangle
\end{align}
with one additional phonon either to the left or to the right of the phonon cloud. Note that for $n=1$, $F_{\sigma,
  l}^{(1,n)}(k,\omega) = F_{\sigma, l}^{(n,1)}(k,\omega)$.

We now apply Dyson's identity again, to obtain the eom for the
generalized GFs. Suppressing the $(k,\omega)$ argument for simplicity, we find:
\begin{align}
  f_{\sigma, l}^{(n)} &=(n f_{s,0}^{(n-1)} + f_{s,0}^{(n+1)} + n
  f_{s,1}^{(n-1)} + f_{s,1}^{(n+1)}) \bar{g}_{-l}^{p,\sigma} \nonumber
  \\ &+(F_{s,-1}^{(1,n)} + F_{s,0}^{(1,n)})\bar{g}_{-l-1}^{p,\sigma}
  +\e{-ika} (F_{s,1}^{(n,1)} + F_{s,2}^{(n,1)}) \nonumber \\ &\times
  \bar{g}_{-l+1}^{p,\sigma} + (n f_{p,0}^{(n-1)} + f_{p,0}^{(n+1)})
  (\bar{g}_{-l}^{s,\sigma} + \bar{g}_{-l+1}^{s,\sigma}) \nonumber
  \\ &+ F_{p,-1}^{(1,n)} (\bar{g}_{-l-1}^{s,\sigma} +
  \bar{g}_{-l}^{s,\sigma}) + \e{-ika}F_{p,1}^{(n,1)}. \nonumber
  \\ &\times (\bar{g}_{-l+1}^{s,\sigma} + \bar{g}_{-l+2}^{s,\sigma})
\label{eq:eom_f}
\end{align}
\begin{align}
  F_{\sigma,l}^{(1,n)} &= (n F_{s,0}^{(1,n-1)} + F_{s,0}^{(1,n+1)} + n
  F_{s,1}^{(1,n-1)} + F_{s,1}^{(1,n+1)}) \bar{g}_{-l}^{p,\sigma}\nonumber
  \\ &+ (f_{s,-1}^{(n)} + f_{s,0}^{(n)})\bar{g}_{-l-1}^{p,\sigma} +
  (nF_{p,0}^{(1,n-1)} + F_{p,0}^{(1,n+1)}) \nonumber \\ &\times
  (\bar{g}_{-l}^{s,\sigma} + \bar{g}_{-l+1}^{s,\sigma}) +
  f_{p,-1}^{(n)} (\bar{g}_{-l-1}^{s,\sigma} +
  \bar{g}_{-l}^{s,\sigma}).
\label{eq:eom_F(1,n)}
\end{align}
\begin{align}
  F_{\sigma,l}^{(n,1)} &= (nF_{s,0}^{(n-1,1)} F_{s,0}^{(n+1,1)} +
  nF_{s,1}^{(n-1,1)} + F_{s,1}^{(n+1,1)}) \bar{g}_{-l}^{p,\sigma}
  \nonumber \\ &+ (\e{ika} f_{s,1}^{(n)} + \e{ika} f_{s,2}^{(n)})
  \bar{g}_{-l+1}^{p,\sigma} \nonumber \\ &+ (n F_{p,0}^{(n-1,1)} +
  F_{p,0}^{(n+1,1)}) (\bar{g}_{-l}^{s,\sigma} +
  \bar{g}_{-l+1}^{s,\sigma}) \nonumber \\ &+ \e{ika} f_{p,1}^{(n)}
  (\bar{g}_{-l+1}^{s,\sigma} +\bar{g}_{-l+2}^{s,\sigma}),
\label{eq:eom_F(n,1)}
\end{align}
where we introduced the free hole real-space GFs:
$$\bar{g}_{l}^{\sigma',\sigma}\equiv \bar{g}_{l}^{\sigma',\sigma}(\omega - n \Omega) = {g\over N}\sum_q \e{i q R_l}
G_{0}^{\sigma' \sigma}(q,\omega-n\Omega).$$
For simplicity its
$(\omega-n\Omega)$ argument was again suppressed in the equations above. In 1D, these free hole
real-space GFs can be obtained analytically, see Appendix B.

The right hand side of these eom only links to generalized GFs with
specific values of $l$. Let us first consider the case of MA$^{(0)}$.
Here we only consider generalized GFs of the type
$f_{\sigma,l}^{(n)}$. By inspection of Eq. (\ref{eq:eom_f}) we find
that the eom only link to $\tilde{f}_{s}^{(n)} =f_{s,0}^{(n)} +
f_{s,1}^{(n)} $ and $\tilde{f}_{p}^{(n)}= f_{p,0}^{(n)}$. Furthermore
since $\bar{g}_{0}^{p,s}+\bar{g}_{-1}^{p,s}=0$ and $g_{0}^{s,p} +
g_{1}^{s,p}=0$ and $\bar{g}_{-l}^{\sigma,\sigma}=\bar{g}_{l}^{\sigma,\sigma}$ (see Appendix B) the eom simplify to:
\begin{align}
  &\tilde{f}_{s}^{(n)} = 2(\bar{g}_{0}^{s,s}+\bar{g}_{1}^{s,s})
  (n\tilde{f}_{p}^{(n-1)} + \tilde{f}_{p}^{(n+1)})
  \\ &\tilde{f}_{p}^{(n)} = \bar{g}_{0}^{p,p} 
  (n\tilde{f}_{s}^{(n-1)} +\tilde{f}_{s}^{(n+1)}).
\end{align}
We
define the column vector $\mathbf{\tilde{v}}_n = (\tilde{f}_{s}^{(n)},
\tilde{f}_{p}^{(n)})^T$ and recast the eom in the form
\begin{align}
  \mathbf{\tilde{v}}_{n} = \alpha_n \mathbf{\tilde{v}}_{n-1} +
  \beta_{n} \mathbf{\tilde{v}}_{n+1},
\end{align}
where $\tilde{\alpha}_n$ and $\tilde{\beta}_n$ are $2\times
2$-matrices which for $n\ge 2$ can be read off directly from the
equations above. To get $\tilde{\alpha}_1$ we use 
$\tilde{f}_{s}^{(0)}=(1+\e{-ika})G^{\sigma's}$ and $
\tilde{f}_{p}^{(0)} = G^{\sigma'p}$, where  $\sigma'$ is the same
as in Eq. (\ref{eq:eom_G}). The eom can now be solved with the ansatz
$\mathbf{\tilde{v}}_{n}=\tilde{A}_{n} \mathbf{\tilde{v}}_{n-1}$,
\cite{Glen-BM} which is justified because for large $n$,
$\bar{g}_{l}^{\sigma',\sigma}(\omega-n\Omega)$ goes to zero and
therefore $\mathbf{\tilde{v}}_{n}$ must go to zero as well. Plugging
this ansatz back into Eq. (\ref{eq:recurrence}) yields
$\tilde{A}_{n}=[1-\tilde{\beta}_{n}\tilde{A}_{n+1}]^{-1}\tilde{\alpha}_n$
which can be calculated recursively starting with $\tilde{A}_{M_c+1}=0$, where
$M_c$ is chosen sufficiently large so that its further increase has no
effect on the results.

Once we have calculated $\tilde{A}_{1}$ linking
$\mathbf{\tilde{v}}_{1}$ to $\mathbf{\tilde{v}}_{0}^{\sigma'} =
(G^{\sigma's}, G^{\sigma'p})^T$, we can rewrite Eq. (\ref{eq:eom_G})
in matrix form:
\begin{align}
  G(k,\omega) = G_0(k,\omega) + g G(k,\omega) \tilde{A}_{1}^T
  \tilde{M} G_0(k,\omega),
  \label{eq:eom_G_matrix_MA0}
\end{align}
Note that this requires identifying the first row of $G(k,\omega)$
with $(\mathbf{v}_{0}^s)^T$ and the second row with
$(\mathbf{v}_{0}^p)^T$. $\tilde{M}$ is the $2 \times 2$-matrix with
$\tilde{M}_{1,1}=\tilde{M}_{2,2}=0, \tilde{M}_{1,2}=1,
\tilde{M}_{2,1}=1+\e{ika}$. The self-energy $\Sigma(k,\omega)$ is
defined by the $2\times 2$-matrix equation $G(k,\omega) =
[G_{0}(k,\omega)-\Sigma(k,\omega)]^{-1}$. In the MA$^{(0)}$
approximation it is therefore given by:
\begin{align}
  {\Sigma}_{(0)}(k,\omega) = g \tilde{A}_{1}^T \tilde{M}.
\end{align}

We now show how to use the MA$^{(0)}$ eom to rigorously derive the
perturbation result of Sec. \ref{sec:perturbation-theory}. For a
cutoff of $M_c=1$, we find that
\begin{align}
  \tilde{A}_1 = \tilde{\alpha_1} = \left (
    \begin{array}{cc}
      0 & 2(\bar{g}_{0}^{s,s} + \bar{g}_{1}^{s,s}) \\
      \bar{g}_{0}^{p,p}(1 + \e{-ika}) & 0 \\
    \end{array} \right )
\end{align}
and consequently
\begin{align}
  \tilde{\Sigma}^{(M_c=1)} = 2g\left (
    \begin{array}{cc}
      (1+\cos(ka))\bar{g}_{0}^{p,p} & 0 
      \\ 0 & \bar{g}_{0}^{s,s} + \bar{g}_{1}^{s,s} \\
    \end{array} \right )
\end{align}
For $\Omega \gg t,\Delta$ we have $\bar{g}_{0}^{s,s}
\approx \bar{g}_{0}^{p,p}\approx -g/\Omega$ and
$\bar{g}_{1}^{s,s}\approx 0$ (see Appendix B) and we recover the
result of Eq. (\ref{eq:Heff_matrix}).
 
The MA$^{(1r)}$ case is treated in
exactly the same manner, but is slightly more tedious. For
$n> 2 $  phonons we include the following 17 GFs:
$f_{s,l}^{(n)}$ with $l \in \{-1,0,1,2 \}$; $f_{p,l}^{(n)}$ with $ l \in
  \{-1,0,1 \}$; $F_{s,l}^{(1,n-1)}$ with $l \in \{-1,0,1\}$;
  $F_{p,l}^{(1,n-1)}$ with $ l \in \{-1,0\}$; $F_{s,l}^{(n-1,1)}$ with $ l \in
  \{0,1,2 \}$; and $F_{p,l}^{(n-1,1)}$ with $l \in \{0,1\}$.
For $n=2$, some of these GFs are identical so that their number is
reduced to 12. Similarly, for $n=1$ one only needs to keep 7 GFs.

The
generalized GFs are again arranged in a vector $\mathbf{v}_{n}$ and
the eom recast as a recurrence equation:
\begin{align}
 \mathbf{v}_{n} = \alpha_n \mathbf{v}_{n-1} + \beta_{n}
 \mathbf{v}_{n+1}.
\label{eq:recurrence}
\end{align}
The matrices $\alpha_n$ and $\beta_n$ can be read off  from
Eqs. (\ref{eq:eom_f}), (\ref{eq:eom_F(1,n)}) and (\ref{eq:eom_F(n,1)}).
Furthermore, Eq. (\ref{eq:def_f}) indicates that $(f_{s,0}^{(0)},
f_{s,1}^{(0)}, f_{p,0}^{0}) = (G^{\sigma's}, \e{-ika} G^{\sigma's},
G^{\sigma'p})$ which we use to read off $\alpha_1$. Again the eom 
are solved with the ansatz $\mathbf{v}_{n}=A_{n} \mathbf{v}_{n-1}$
\cite{Glen-BM}, yielding $A_{n}=[1-\beta_{n}A_{n+1}]^{-1}\alpha_n$.

We then rewrite Eq. (\ref{eq:eom_G}) in matrix form:
\begin{align}
  G(k,\omega) = G_0(k,\omega) + g G(k,\omega) A_{1,r}^T M
  G_0(k,\omega),
  \label{eq:eom_G_matrix}
\end{align}
Here $A_{1,r}$\ is a reduced version of $A_{1}$. It is a $3\times
2$-matrix which only contains the rows of $A_{1}$ linking
$f_{s,0}^{(1)}, f_{s,1}^{(1)}$ and $f_{p,0}^{(1)}$ to
$\mathbf{v}_{0}$. This is necessary since the other 4 generalized GFs
contained in $\mathbf{v}_{1}$ do not appear in Eq. (\ref{eq:eom_G}).
The matrix $M$ is a $3 \times 2$-matrix whose only non-zero elements
are $M_{3,1}=1+\e{ika}$ and $ M_{1,2}=M_{2,2}=1$. The self-energy
$\Sigma(k,\omega)$ in the MA$^{(1r)}$ approximation is, then:
\begin{align}
  \Sigma_{(1r)}(k,\omega) = g \tilde{A}_{1,r}^T M.
\end{align}

\section{Real-Space Green's functions (GFs)}
The real-space GFs of $\hat{H}_0=\hat{T} + \hat{H}_{\text{ct}} +
\hat{H}_{\text{ph}}$ are defined as
$g_{0}^{\sigma',\sigma}(R_l,\omega) = \sum_q
\frac{\e{iqR_{j+l}}}{\sqrt{N}} G_0(q,\omega)$. This can be rewritten
as $ g_{0}^{\sigma',\sigma}(R_l,\omega) = \langle 0 | c_{\sigma',l}
\hat{G}(\omega) c_{\sigma,0} | 0 \rangle$, {\em i.e.}
$g_{0}^{\sigma',\sigma}(R_l,\omega)$ measures the probability
amplitude that a hole injected at site 0 will be removed at site $l$.
The eom are obtained using Dyson's identity. For $\sigma'=p$, and
suppressing the $(k,\omega)$-dependence, we find:
\begin{align}
  &g_{0}^{p,p}(\tilde{\omega}+\Delta) = 1 - t g_{-1}^{p,s} + t
  g_{0}^{p,s} \\ &g_{-1}^{p,s}\tilde{\omega} = -t g_{0}^{p,p} + t
  g_{-1}^{p,p} \\ &g_{0}^{p,s}\tilde{\omega} = -t g_{1}^{p,p} + t
  g_{0}^{p,p},
\end{align}
where we defined the shorthand $\tilde{\omega}=\omega+i\eta$. For $n
\neq 0,-1$, the general form of the eom is
\begin{align}
  &g_{n}^{p,p}(\tilde{\omega}+\Delta) = -t g_{n-1}^{p,s} + t
  g_{n}^{p,s} \\ &g_{n}^{p,s}\tilde{\omega} = -t g_{n+1}^{p,p} + t
  g_{1}^{p,p}
\end{align}
Eliminating $g_{n}^{p,s}$, the eom for $g_{n}^{p,p}$, with $n \neq
0,-1$, can be recast as
\begin{align}
  g_{n}^{p,p}[\tilde{\omega}(\tilde{\omega}+\Delta)-2t^2] = -t^2
  g_{n-1}^{p,p} - t^2 g_{n+1}^{p,p}
\end{align}
Similarly we find
\begin{align}
  g_{0}^{p,p}(\tilde{\omega}(\tilde{\omega}+\Delta)-2t^2)=\tilde{\omega}
  - t^2 (g_{-1}^{p,p} + g_{1}^{p,p})
  \label{eq:eom_g0pp}
\end{align}
In the time-domain the small imaginary part $i\eta$ corresponds to a
finite lifetime of the hole. Therefore the probability that a hole
travels from site 0 to site $j$ falls off exponentially as $\eta j$.
and we use the ansatz $g_n^{p,p} = z g_{n-1}^{p,p}$, for $n>0$. For
$n<-1$ we need to use $g_n^{p,p} = z g_{n+1}^{p,p}$. In both cases,
plugging the ansatz back into the eom gives:
\begin{align}
  z_{\pm} = - \left( \frac{\tilde{\omega}(\tilde{\omega}+\Delta)}
  {2t^2}-1 \right) \pm \sqrt{\left(
    \frac{\tilde{\omega}(\tilde{\omega}+\Delta)} {2t^2}-1
    \right)^2-1},
  \label{eq:z}
\end{align}
We need to choose the solution which satisfies $|z|<1$. Using the
ansatz in Eq. (\ref{eq:eom_g0pp}) we obtain
\begin{align}
  g_0^{p,p} = \frac{\tilde{\omega}}
  {\tilde{\omega}(\tilde{\omega}+\Delta)+2t^2(z-1)}
\end{align}
From this all the other $g_{n}^{p,p}$ are obtained as $g_{n}^{p,p} =
z^{|n|}g_{0}^{p,p}$.

We can apply the same procedure to find analytical expressions for the
$g_{n}^{p,s}$. However, we need to be careful since both
$g_{-1}^{p,s}$ and $g_{0}^{p,s}$ link to $g_{0}^{pp}$ and therefore
need to be treated separately. After some algebra we find
\begin{align}
  g_{0}^{p,s} = -g_{-1}^{p,s} =
  \frac{t}{\tilde{\omega}(\tilde{\omega}+\Delta)+t^2(z-3)}
\end{align}
\begin{align}
  &g_{n}^{p,s} = z^n g_{0}^{p,s}, &n>0 \\ &g_{n}^{p,s} = z^{|n|-1}
  g_{-1}^{p,s}, &n<-1
\end{align}
Similarly one finds:
\begin{align}
  &g_0^{s,s} =\frac{\tilde{\omega}+\Delta}
  {\tilde{\omega}(\tilde{\omega}+\Delta)+2t^2(z-1)} \\ &g_{n}^{s,s} =
  z^{|n|}g_{0}^{s,s} \\ &g_{0}^{s,p} = -g_{1}^{s,p} =
  \frac{t}{\tilde{\omega}(\tilde{\omega}+\Delta)+t^2(z-3)}
\end{align}
\begin{align}
  &g_{n}^{p,s} = z^{|n|} g_{0}^{p,s}, &n<0 \\ &g_{n}^{p,s} = z^{n-1}
  g_{1}^{p,s}, &n>1
\end{align}

Note that for $|\omega| \gg t, \Delta$, Eq. (\ref{eq:z}) implies that
$z\rightarrow 0$. Using this in the expressions for
$g_{0,-1}^{p,s}(\omega)$ and $g_{0,1}^{s,p}(\omega)$ we find that they
scale as $\pm 1/\omega^2$. The diagonal real-space GFs, $g_{0}^{s,s}$
and $g_{0}^{p,p}$ on the other hand scale as $1/\omega$. All
real-space GFs with larger values of $|n|$ go to 0 since $z
\rightarrow 0$.

% \bibliography{article04.bib}

\begin{thebibliography}{54}%
\makeatletter
\providecommand \@ifxundefined [1]{%
 \@ifx{#1\undefined}
}%
\providecommand \@ifnum [1]{%
 \ifnum #1\expandafter \@firstoftwo
 \else \expandafter \@secondoftwo
 \fi
}%
\providecommand \@ifx [1]{%
 \ifx #1\expandafter \@firstoftwo
 \else \expandafter \@secondoftwo
 \fi
}%
\providecommand \natexlab [1]{#1}%
\providecommand \enquote  [1]{``#1''}%
\providecommand \bibnamefont  [1]{#1}%
\providecommand \bibfnamefont [1]{#1}%
\providecommand \citenamefont [1]{#1}%
\providecommand \href@noop [0]{\@secondoftwo}%
\providecommand \href [0]{\begingroup \@sanitize@url \@href}%
\providecommand \@href[1]{\@@startlink{#1}\@@href}%
\providecommand \@@href[1]{\endgroup#1\@@endlink}%
\providecommand \@sanitize@url [0]{\catcode `\\12\catcode `\$12\catcode
  `\&12\catcode `\#12\catcode `\^12\catcode `\_12\catcode `\%12\relax}%
\providecommand \@@startlink[1]{}%
\providecommand \@@endlink[0]{}%
\providecommand \url  [0]{\begingroup\@sanitize@url \@url }%
\providecommand \@url [1]{\endgroup\@href {#1}{\urlprefix }}%
\providecommand \urlprefix  [0]{URL }%
\providecommand \Eprint [0]{\href }%
\providecommand \doibase [0]{http://dx.doi.org/}%
\providecommand \selectlanguage [0]{\@gobble}%
\providecommand \bibinfo  [0]{\@secondoftwo}%
\providecommand \bibfield  [0]{\@secondoftwo}%
\providecommand \translation [1]{[#1]}%
\providecommand \BibitemOpen [0]{}%
\providecommand \bibitemStop [0]{}%
\providecommand \bibitemNoStop [0]{.\EOS\space}%
\providecommand \EOS [0]{\spacefactor3000\relax}%
\providecommand \BibitemShut  [1]{\csname bibitem#1\endcsname}%
\let\auto@bib@innerbib\@empty
%</preamble>
\bibitem [{\citenamefont {R\"osch}\ and\ \citenamefont
  {Gunnarsson}(2004)}]{cuprates}%
  \BibitemOpen
  \bibfield  {author} {\bibinfo {author} {\bibfnamefont {O.}~\bibnamefont
  {R\"osch}}\ and\ \bibinfo {author} {\bibfnamefont {O.}~\bibnamefont
  {Gunnarsson}},\ }\href {\doibase 10.1103/PhysRevLett.92.146403} {\bibfield
  {journal} {\bibinfo  {journal} {Phys. Rev. Lett.}\ }\textbf {\bibinfo
  {volume} {92}},\ \bibinfo {pages} {146403} (\bibinfo {year}
  {2004})}\BibitemShut {NoStop}%
\bibitem [{\citenamefont {M{\"u}ller}(1999)}]{Jahn-Teller-Polaron}%
  \BibitemOpen
  \bibfield  {author} {\bibinfo {author} {\bibfnamefont {K.}~\bibnamefont
  {M{\"u}ller}},\ }\href {\doibase 10.1023/A:1007792713484} {\bibfield
  {journal} {\bibinfo  {journal} {Journal of Superconductivity}\ }\textbf
  {\bibinfo {volume} {12}},\ \bibinfo {pages} {3} (\bibinfo {year}
  {1999})}\BibitemShut {NoStop}%
\bibitem [{\citenamefont {Driza}\ \emph {et~al.}(2012)\citenamefont {Driza},
  \citenamefont {Blanco-Canosa}, \citenamefont {Bakr}, \citenamefont {Soltan},
  \citenamefont {Khalid}, \citenamefont {Mustafa}, \citenamefont {Kawashima},
  \citenamefont {Christiani}, \citenamefont {Habermeier}, \citenamefont
  {Khaliullin}, \citenamefont {Ulrich}, \citenamefont {Le~Tacon},\ and\
  \citenamefont {Keimer}}]{manganites+cuprates}%
  \BibitemOpen
  \bibfield  {author} {\bibinfo {author} {\bibfnamefont {N.}~\bibnamefont
  {Driza}}, \bibinfo {author} {\bibfnamefont {S.}~\bibnamefont
  {Blanco-Canosa}}, \bibinfo {author} {\bibfnamefont {M.}~\bibnamefont {Bakr}},
  \bibinfo {author} {\bibfnamefont {S.}~\bibnamefont {Soltan}}, \bibinfo
  {author} {\bibfnamefont {M.}~\bibnamefont {Khalid}}, \bibinfo {author}
  {\bibfnamefont {L.}~\bibnamefont {Mustafa}}, \bibinfo {author} {\bibfnamefont
  {K.}~\bibnamefont {Kawashima}}, \bibinfo {author} {\bibfnamefont
  {G.}~\bibnamefont {Christiani}}, \bibinfo {author} {\bibfnamefont {H.-U.}\
  \bibnamefont {Habermeier}}, \bibinfo {author} {\bibfnamefont
  {G.}~\bibnamefont {Khaliullin}}, \bibinfo {author} {\bibfnamefont
  {C.}~\bibnamefont {Ulrich}}, \bibinfo {author} {\bibfnamefont
  {M.}~\bibnamefont {Le~Tacon}}, \ and\ \bibinfo {author} {\bibfnamefont
  {B.}~\bibnamefont {Keimer}},\ }\href {\doibase 10.1038/nmat3378} {\bibfield
  {journal} {\bibinfo  {journal} {Nat Mater}\ }\textbf {\bibinfo {volume}
  {11}},\ \bibinfo {pages} {675} (\bibinfo {year} {2012})}\BibitemShut
  {NoStop}%
\bibitem [{\citenamefont {Gunnarsson}\ and\ \citenamefont
  {Rösch}(2008)}]{cuprates-phonons-review}%
  \BibitemOpen
  \bibfield  {author} {\bibinfo {author} {\bibfnamefont {O.}~\bibnamefont
  {Gunnarsson}}\ and\ \bibinfo {author} {\bibfnamefont {O.}~\bibnamefont
  {Rösch}},\ }\href {http://stacks.iop.org/0953-8984/20/i=4/a=043201}
  {\bibfield  {journal} {\bibinfo  {journal} {Journal of Physics: Condensed
  Matter}\ }\textbf {\bibinfo {volume} {20}},\ \bibinfo {pages} {043201}
  (\bibinfo {year} {2008})}\BibitemShut {NoStop}%
\bibitem [{\citenamefont {Edwards}(2002)}]{manganites-phonons}%
  \BibitemOpen
  \bibfield  {author} {\bibinfo {author} {\bibfnamefont {D.~M.}\ \bibnamefont
  {Edwards}},\ }\href {\doibase 10.1080/00018730210140805} {\bibfield
  {journal} {\bibinfo  {journal} {Advances in Physics}\ }\textbf {\bibinfo
  {volume} {51}},\ \bibinfo {pages} {1259} (\bibinfo {year}
  {2002})}\BibitemShut {NoStop}%
\bibitem [{\citenamefont {Johnston}\ \emph {et~al.}(2014)\citenamefont
  {Johnston}, \citenamefont {Mukherjee}, \citenamefont {Elfimov}, \citenamefont
  {Berciu},\ and\ \citenamefont {Sawatzky}}]{Steve-nickelates}%
  \BibitemOpen
  \bibfield  {author} {\bibinfo {author} {\bibfnamefont {S.}~\bibnamefont
  {Johnston}}, \bibinfo {author} {\bibfnamefont {A.}~\bibnamefont {Mukherjee}},
  \bibinfo {author} {\bibfnamefont {I.}~\bibnamefont {Elfimov}}, \bibinfo
  {author} {\bibfnamefont {M.}~\bibnamefont {Berciu}}, \ and\ \bibinfo {author}
  {\bibfnamefont {G.~A.}\ \bibnamefont {Sawatzky}},\ }\href {\doibase
  10.1103/PhysRevLett.112.106404} {\bibfield  {journal} {\bibinfo  {journal}
  {Phys. Rev. Lett.}\ }\textbf {\bibinfo {volume} {112}},\ \bibinfo {pages}
  {106404} (\bibinfo {year} {2014})}\BibitemShut {NoStop}%
\bibitem [{\citenamefont {Gou}\ \emph {et~al.}(2011)\citenamefont {Gou},
  \citenamefont {Grinberg}, \citenamefont {Rappe},\ and\ \citenamefont
  {Rondinelli}}]{nickelates-normal-modes}%
  \BibitemOpen
  \bibfield  {author} {\bibinfo {author} {\bibfnamefont {G.}~\bibnamefont
  {Gou}}, \bibinfo {author} {\bibfnamefont {I.}~\bibnamefont {Grinberg}},
  \bibinfo {author} {\bibfnamefont {A.~M.}\ \bibnamefont {Rappe}}, \ and\
  \bibinfo {author} {\bibfnamefont {J.~M.}\ \bibnamefont {Rondinelli}},\ }\href
  {\doibase 10.1103/PhysRevB.84.144101} {\bibfield  {journal} {\bibinfo
  {journal} {Phys. Rev. B}\ }\textbf {\bibinfo {volume} {84}},\ \bibinfo
  {pages} {144101} (\bibinfo {year} {2011})}\BibitemShut {NoStop}%
\bibitem [{\citenamefont {Zaanen}\ and\ \citenamefont
  {Littlewood}(1994)}]{nickelates-Zaanen}%
  \BibitemOpen
  \bibfield  {author} {\bibinfo {author} {\bibfnamefont {J.}~\bibnamefont
  {Zaanen}}\ and\ \bibinfo {author} {\bibfnamefont {P.~B.}\ \bibnamefont
  {Littlewood}},\ }\href {\doibase 10.1103/PhysRevB.50.7222} {\bibfield
  {journal} {\bibinfo  {journal} {Phys. Rev. B}\ }\textbf {\bibinfo {volume}
  {50}},\ \bibinfo {pages} {7222} (\bibinfo {year} {1994})}\BibitemShut
  {NoStop}%
\bibitem [{\citenamefont {Foyevtsova}\ \emph {et~al.}(2015)\citenamefont
  {Foyevtsova}, \citenamefont {Khazraie}, \citenamefont {Elfimov},\ and\
  \citenamefont {Sawatzky}}]{Kateryna}%
  \BibitemOpen
  \bibfield  {author} {\bibinfo {author} {\bibfnamefont {K.}~\bibnamefont
  {Foyevtsova}}, \bibinfo {author} {\bibfnamefont {A.}~\bibnamefont
  {Khazraie}}, \bibinfo {author} {\bibfnamefont {I.}~\bibnamefont {Elfimov}}, \
  and\ \bibinfo {author} {\bibfnamefont {G.~A.}\ \bibnamefont {Sawatzky}},\
  }\href {\doibase 10.1103/PhysRevB.91.121114} {\bibfield  {journal} {\bibinfo
  {journal} {Phys. Rev. B}\ }\textbf {\bibinfo {volume} {91}},\ \bibinfo
  {pages} {121114} (\bibinfo {year} {2015})}\BibitemShut {NoStop}%
\bibitem [{\citenamefont {Franchini}\ \emph {et~al.}(2009)\citenamefont
  {Franchini}, \citenamefont {Kresse},\ and\ \citenamefont
  {Podloucky}}]{Franchini-BaBiO3}%
  \BibitemOpen
  \bibfield  {author} {\bibinfo {author} {\bibfnamefont {C.}~\bibnamefont
  {Franchini}}, \bibinfo {author} {\bibfnamefont {G.}~\bibnamefont {Kresse}}, \
  and\ \bibinfo {author} {\bibfnamefont {R.}~\bibnamefont {Podloucky}},\ }\href
  {\doibase 10.1103/PhysRevLett.102.256402} {\bibfield  {journal} {\bibinfo
  {journal} {Phys. Rev. Lett.}\ }\textbf {\bibinfo {volume} {102}},\ \bibinfo
  {pages} {256402} (\bibinfo {year} {2009})}\BibitemShut {NoStop}%
\bibitem [{\citenamefont {Bon\ifmmode~\check{c}\else \v{c}\fi{}a}\ \emph
  {et~al.}(1999)\citenamefont {Bon\ifmmode~\check{c}\else \v{c}\fi{}a},
  \citenamefont {Trugman},\ and\ \citenamefont {Batisti\ifmmode~\acute{c}\else
  \'{c}\fi{}}}]{Bonca+Trugman1}%
  \BibitemOpen
  \bibfield  {author} {\bibinfo {author} {\bibfnamefont {J.}~\bibnamefont
  {Bon\ifmmode~\check{c}\else \v{c}\fi{}a}}, \bibinfo {author} {\bibfnamefont
  {S.~A.}\ \bibnamefont {Trugman}}, \ and\ \bibinfo {author} {\bibfnamefont
  {I.}~\bibnamefont {Batisti\ifmmode~\acute{c}\else \'{c}\fi{}}},\ }\href
  {\doibase 10.1103/PhysRevB.60.1633} {\bibfield  {journal} {\bibinfo
  {journal} {Phys. Rev. B}\ }\textbf {\bibinfo {volume} {60}},\ \bibinfo
  {pages} {1633} (\bibinfo {year} {1999})}\BibitemShut {NoStop}%
\bibitem [{\citenamefont {Ku}\ \emph {et~al.}(2002)\citenamefont {Ku},
  \citenamefont {Trugman},\ and\ \citenamefont {Bon\ifmmode~\check{c}\else
  \v{c}\fi{}a}}]{Bonca+Trugman2}%
  \BibitemOpen
  \bibfield  {author} {\bibinfo {author} {\bibfnamefont {L.-C.}\ \bibnamefont
  {Ku}}, \bibinfo {author} {\bibfnamefont {S.~A.}\ \bibnamefont {Trugman}}, \
  and\ \bibinfo {author} {\bibfnamefont {J.}~\bibnamefont
  {Bon\ifmmode~\check{c}\else \v{c}\fi{}a}},\ }\href {\doibase
  10.1103/PhysRevB.65.174306} {\bibfield  {journal} {\bibinfo  {journal} {Phys.
  Rev. B}\ }\textbf {\bibinfo {volume} {65}},\ \bibinfo {pages} {174306}
  (\bibinfo {year} {2002})}\BibitemShut {NoStop}%
\bibitem [{\citenamefont {Vidmar}\ \emph {et~al.}(2010)\citenamefont {Vidmar},
  \citenamefont {Bon\ifmmode~\check{c}\else \v{c}\fi{}a},\ and\ \citenamefont
  {Trugman}}]{Bonca+Trugman3}%
  \BibitemOpen
  \bibfield  {author} {\bibinfo {author} {\bibfnamefont {L.}~\bibnamefont
  {Vidmar}}, \bibinfo {author} {\bibfnamefont {J.}~\bibnamefont
  {Bon\ifmmode~\check{c}\else \v{c}\fi{}a}}, \ and\ \bibinfo {author}
  {\bibfnamefont {S.~A.}\ \bibnamefont {Trugman}},\ }\href {\doibase
  10.1103/PhysRevB.82.104304} {\bibfield  {journal} {\bibinfo  {journal} {Phys.
  Rev. B}\ }\textbf {\bibinfo {volume} {82}},\ \bibinfo {pages} {104304}
  (\bibinfo {year} {2010})}\BibitemShut {NoStop}%
\bibitem [{\citenamefont {Li}\ \emph {et~al.}(2010)\citenamefont {Li},
  \citenamefont {Baillie}, \citenamefont {Blois},\ and\ \citenamefont
  {Marsiglio}}]{Li}%
  \BibitemOpen
  \bibfield  {author} {\bibinfo {author} {\bibfnamefont {Z.}~\bibnamefont
  {Li}}, \bibinfo {author} {\bibfnamefont {D.}~\bibnamefont {Baillie}},
  \bibinfo {author} {\bibfnamefont {C.}~\bibnamefont {Blois}}, \ and\ \bibinfo
  {author} {\bibfnamefont {F.}~\bibnamefont {Marsiglio}},\ }\href {\doibase
  10.1103/PhysRevB.81.115114} {\bibfield  {journal} {\bibinfo  {journal} {Phys.
  Rev. B}\ }\textbf {\bibinfo {volume} {81}},\ \bibinfo {pages} {115114}
  (\bibinfo {year} {2010})}\BibitemShut {NoStop}%
\bibitem [{\citenamefont {Alvermann}\ \emph {et~al.}(2010)\citenamefont
  {Alvermann}, \citenamefont {Fehske},\ and\ \citenamefont
  {Trugman}}]{Alvermann+Trugman}%
  \BibitemOpen
  \bibfield  {author} {\bibinfo {author} {\bibfnamefont {A.}~\bibnamefont
  {Alvermann}}, \bibinfo {author} {\bibfnamefont {H.}~\bibnamefont {Fehske}}, \
  and\ \bibinfo {author} {\bibfnamefont {S.~A.}\ \bibnamefont {Trugman}},\
  }\href {\doibase 10.1103/PhysRevB.81.165113} {\bibfield  {journal} {\bibinfo
  {journal} {Phys. Rev. B}\ }\textbf {\bibinfo {volume} {81}},\ \bibinfo
  {pages} {165113} (\bibinfo {year} {2010})}\BibitemShut {NoStop}%
\bibitem [{\citenamefont {Chandler}\ and\ \citenamefont
  {Marsiglio}(2014)}]{Holstein-Long-Range}%
  \BibitemOpen
  \bibfield  {author} {\bibinfo {author} {\bibfnamefont {C.~J.}\ \bibnamefont
  {Chandler}}\ and\ \bibinfo {author} {\bibfnamefont {F.}~\bibnamefont
  {Marsiglio}},\ }\href {\doibase 10.1103/PhysRevB.90.125131} {\bibfield
  {journal} {\bibinfo  {journal} {Phys. Rev. B}\ }\textbf {\bibinfo {volume}
  {90}},\ \bibinfo {pages} {125131} (\bibinfo {year} {2014})}\BibitemShut
  {NoStop}%
\bibitem [{\citenamefont {Berciu}(2006)}]{Mona-MA-1}%
  \BibitemOpen
  \bibfield  {author} {\bibinfo {author} {\bibfnamefont {M.}~\bibnamefont
  {Berciu}},\ }\href {\doibase 10.1103/PhysRevLett.97.036402} {\bibfield
  {journal} {\bibinfo  {journal} {Phys. Rev. Lett.}\ }\textbf {\bibinfo
  {volume} {97}},\ \bibinfo {pages} {036402} (\bibinfo {year}
  {2006})}\BibitemShut {NoStop}%
\bibitem [{\citenamefont {Berciu}\ and\ \citenamefont
  {Goodvin}(2007)}]{Mona-MA-2}%
  \BibitemOpen
  \bibfield  {author} {\bibinfo {author} {\bibfnamefont {M.}~\bibnamefont
  {Berciu}}\ and\ \bibinfo {author} {\bibfnamefont {G.~L.}\ \bibnamefont
  {Goodvin}},\ }\href {\doibase 10.1103/PhysRevB.76.165109} {\bibfield
  {journal} {\bibinfo  {journal} {Phys. Rev. B}\ }\textbf {\bibinfo {volume}
  {76}},\ \bibinfo {pages} {165109} (\bibinfo {year} {2007})}\BibitemShut
  {NoStop}%
\bibitem [{\citenamefont {Lau}\ \emph {et~al.}(2007)\citenamefont {Lau},
  \citenamefont {Berciu},\ and\ \citenamefont {Sawatzky}}]{Bayo}%
  \BibitemOpen
  \bibfield  {author} {\bibinfo {author} {\bibfnamefont {B.}~\bibnamefont
  {Lau}}, \bibinfo {author} {\bibfnamefont {M.}~\bibnamefont {Berciu}}, \ and\
  \bibinfo {author} {\bibfnamefont {G.~A.}\ \bibnamefont {Sawatzky}},\ }\href
  {\doibase 10.1103/PhysRevB.76.174305} {\bibfield  {journal} {\bibinfo
  {journal} {Phys. Rev. B}\ }\textbf {\bibinfo {volume} {76}},\ \bibinfo
  {pages} {174305} (\bibinfo {year} {2007})}\BibitemShut {NoStop}%
\bibitem [{\citenamefont {Pankaj}\ and\ \citenamefont
  {Yarlagadda}(2012)}]{CBM}%
  \BibitemOpen
  \bibfield  {author} {\bibinfo {author} {\bibfnamefont {R.}~\bibnamefont
  {Pankaj}}\ and\ \bibinfo {author} {\bibfnamefont {S.}~\bibnamefont
  {Yarlagadda}},\ }\href {\doibase 10.1103/PhysRevB.86.035453} {\bibfield
  {journal} {\bibinfo  {journal} {Phys. Rev. B}\ }\textbf {\bibinfo {volume}
  {86}},\ \bibinfo {pages} {035453} (\bibinfo {year} {2012})}\BibitemShut
  {NoStop}%
\bibitem [{\citenamefont {Goodvin}\ and\ \citenamefont
  {Berciu}(2008)}]{Glen-BM}%
  \BibitemOpen
  \bibfield  {author} {\bibinfo {author} {\bibfnamefont {G.~L.}\ \bibnamefont
  {Goodvin}}\ and\ \bibinfo {author} {\bibfnamefont {M.}~\bibnamefont
  {Berciu}},\ }\href {\doibase 10.1103/PhysRevB.78.235120} {\bibfield
  {journal} {\bibinfo  {journal} {Phys. Rev. B}\ }\textbf {\bibinfo {volume}
  {78}},\ \bibinfo {pages} {235120} (\bibinfo {year} {2008})}\BibitemShut
  {NoStop}%
\bibitem [{\citenamefont {Adolphs}\ and\ \citenamefont
  {Berciu}(2014{\natexlab{a}})}]{Clemens1}%
  \BibitemOpen
  \bibfield  {author} {\bibinfo {author} {\bibfnamefont {C.~P.~J.}\
  \bibnamefont {Adolphs}}\ and\ \bibinfo {author} {\bibfnamefont
  {M.}~\bibnamefont {Berciu}},\ }\href {\doibase 10.1103/PhysRevB.89.035122}
  {\bibfield  {journal} {\bibinfo  {journal} {Phys. Rev. B}\ }\textbf {\bibinfo
  {volume} {89}},\ \bibinfo {pages} {035122} (\bibinfo {year}
  {2014}{\natexlab{a}})}\BibitemShut {NoStop}%
\bibitem [{\citenamefont {Adolphs}\ and\ \citenamefont
  {Berciu}(2014{\natexlab{b}})}]{Clemens2}%
  \BibitemOpen
  \bibfield  {author} {\bibinfo {author} {\bibfnamefont {C.~P.~J.}\
  \bibnamefont {Adolphs}}\ and\ \bibinfo {author} {\bibfnamefont
  {M.}~\bibnamefont {Berciu}},\ }\href {\doibase 10.1103/PhysRevB.90.085149}
  {\bibfield  {journal} {\bibinfo  {journal} {Phys. Rev. B}\ }\textbf {\bibinfo
  {volume} {90}},\ \bibinfo {pages} {085149} (\bibinfo {year}
  {2014}{\natexlab{b}})}\BibitemShut {NoStop}%
\bibitem [{\citenamefont {Fr\"ohlich}\ \emph {et~al.}(1950)\citenamefont
  {Fr\"ohlich}, \citenamefont {Pelzer},\ and\ \citenamefont
  {Zienau}}]{Froehlich1}%
  \BibitemOpen
  \bibfield  {author} {\bibinfo {author} {\bibfnamefont {H.}~\bibnamefont
  {Fr\"ohlich}}, \bibinfo {author} {\bibfnamefont {H.}~\bibnamefont {Pelzer}},
  \ and\ \bibinfo {author} {\bibfnamefont {S.}~\bibnamefont {Zienau}},\ }\href
  {\doibase 10.1080/14786445008521794} {\bibfield  {journal} {\bibinfo
  {journal} {The London, Edinburgh, and Dublin Philosophical Magazine and
  Journal of Science}\ }\textbf {\bibinfo {volume} {41}},\ \bibinfo {pages}
  {221} (\bibinfo {year} {1950})}\BibitemShut {NoStop}%
\bibitem [{\citenamefont {Fr\"ohlich}(1954)}]{Froehlich2}%
  \BibitemOpen
  \bibfield  {author} {\bibinfo {author} {\bibfnamefont {H.}~\bibnamefont
  {Fr\"ohlich}},\ }\href {\doibase 10.1080/00018735400101213} {\bibfield
  {journal} {\bibinfo  {journal} {Advances in Physics}\ }\textbf {\bibinfo
  {volume} {3}},\ \bibinfo {pages} {325} (\bibinfo {year} {1954})}\BibitemShut
  {NoStop}%
\bibitem [{\citenamefont {Su}\ \emph {et~al.}(1979)\citenamefont {Su},
  \citenamefont {Schrieffer},\ and\ \citenamefont
  {Heeger}}]{SSH-original-paper}%
  \BibitemOpen
  \bibfield  {author} {\bibinfo {author} {\bibfnamefont {W.~P.}\ \bibnamefont
  {Su}}, \bibinfo {author} {\bibfnamefont {J.~R.}\ \bibnamefont {Schrieffer}},
  \ and\ \bibinfo {author} {\bibfnamefont {A.~J.}\ \bibnamefont {Heeger}},\
  }\href {\doibase 10.1103/PhysRevLett.42.1698} {\bibfield  {journal} {\bibinfo
   {journal} {Phys. Rev. Lett.}\ }\textbf {\bibinfo {volume} {42}},\ \bibinfo
  {pages} {1698} (\bibinfo {year} {1979})}\BibitemShut {NoStop}%
\bibitem [{\citenamefont {Marchand}\ \emph {et~al.}(2010)\citenamefont
  {Marchand}, \citenamefont {De~Filippis}, \citenamefont {Cataudella},
  \citenamefont {Berciu}, \citenamefont {Nagaosa}, \citenamefont {Prokof'ev},
  \citenamefont {Mishchenko},\ and\ \citenamefont {Stamp}}]{DominicPRL}%
  \BibitemOpen
  \bibfield  {author} {\bibinfo {author} {\bibfnamefont {D.~J.~J.}\
  \bibnamefont {Marchand}}, \bibinfo {author} {\bibfnamefont {G.}~\bibnamefont
  {De~Filippis}}, \bibinfo {author} {\bibfnamefont {V.}~\bibnamefont
  {Cataudella}}, \bibinfo {author} {\bibfnamefont {M.}~\bibnamefont {Berciu}},
  \bibinfo {author} {\bibfnamefont {N.}~\bibnamefont {Nagaosa}}, \bibinfo
  {author} {\bibfnamefont {N.~V.}\ \bibnamefont {Prokof'ev}}, \bibinfo {author}
  {\bibfnamefont {A.~S.}\ \bibnamefont {Mishchenko}}, \ and\ \bibinfo {author}
  {\bibfnamefont {P.~C.~E.}\ \bibnamefont {Stamp}},\ }\href {\doibase
  10.1103/PhysRevLett.105.266605} {\bibfield  {journal} {\bibinfo  {journal}
  {Phys. Rev. Lett.}\ }\textbf {\bibinfo {volume} {105}},\ \bibinfo {pages}
  {266605} (\bibinfo {year} {2010})}\BibitemShut {NoStop}%
\bibitem [{\citenamefont {Zhang}\ \emph {et~al.}(2012)\citenamefont {Zhang},
  \citenamefont {Duan}, \citenamefont {Chen},\ and\ \citenamefont
  {Zhao}}]{Zhao}%
  \BibitemOpen
  \bibfield  {author} {\bibinfo {author} {\bibfnamefont {Y.}~\bibnamefont
  {Zhang}}, \bibinfo {author} {\bibfnamefont {L.}~\bibnamefont {Duan}},
  \bibinfo {author} {\bibfnamefont {Q.}~\bibnamefont {Chen}}, \ and\ \bibinfo
  {author} {\bibfnamefont {Y.}~\bibnamefont {Zhao}},\ }\href {\doibase
  10.1063/1.4733986} {\bibfield  {journal} {\bibinfo  {journal} {The Journal of
  Chemical Physics}\ }\textbf {\bibinfo {volume} {137}},\ \bibinfo {eid}
  {034108} (\bibinfo {year} {2012})}\BibitemShut {NoStop}%
\bibitem [{\citenamefont {Herrera}\ \emph {et~al.}(2013)\citenamefont
  {Herrera}, \citenamefont {Madison}, \citenamefont {Krems},\ and\
  \citenamefont {Berciu}}]{MonaPRL110}%
  \BibitemOpen
  \bibfield  {author} {\bibinfo {author} {\bibfnamefont {F.}~\bibnamefont
  {Herrera}}, \bibinfo {author} {\bibfnamefont {K.~W.}\ \bibnamefont
  {Madison}}, \bibinfo {author} {\bibfnamefont {R.~V.}\ \bibnamefont {Krems}},
  \ and\ \bibinfo {author} {\bibfnamefont {M.}~\bibnamefont {Berciu}},\ }\href
  {\doibase 10.1103/PhysRevLett.110.223002} {\bibfield  {journal} {\bibinfo
  {journal} {Phys. Rev. Lett.}\ }\textbf {\bibinfo {volume} {110}},\ \bibinfo
  {pages} {223002} (\bibinfo {year} {2013})}\BibitemShut {NoStop}%
\bibitem [{\citenamefont {Gerlach}\ and\ \citenamefont
  {L\"owen}(1991)}]{Gerlach}%
  \BibitemOpen
  \bibfield  {author} {\bibinfo {author} {\bibfnamefont {B.}~\bibnamefont
  {Gerlach}}\ and\ \bibinfo {author} {\bibfnamefont {H.}~\bibnamefont
  {L\"owen}},\ }\href {\doibase 10.1103/RevModPhys.63.63} {\bibfield  {journal}
  {\bibinfo  {journal} {Rev. Mod. Phys.}\ }\textbf {\bibinfo {volume} {63}},\
  \bibinfo {pages} {63} (\bibinfo {year} {1991})}\BibitemShut {NoStop}%
\bibitem [{\citenamefont {Cava}\ \emph {et~al.}(1988)\citenamefont {Cava},
  \citenamefont {Batlogg}, \citenamefont {Krajewski}, \citenamefont {Farrow},
  \citenamefont {Rupp}, \citenamefont {White}, \citenamefont {Short},
  \citenamefont {Peck},\ and\ \citenamefont {Kometani}}]{BaKBiO3-Tc}%
  \BibitemOpen
  \bibfield  {author} {\bibinfo {author} {\bibfnamefont {R.~J.}\ \bibnamefont
  {Cava}}, \bibinfo {author} {\bibfnamefont {B.}~\bibnamefont {Batlogg}},
  \bibinfo {author} {\bibfnamefont {J.~J.}\ \bibnamefont {Krajewski}}, \bibinfo
  {author} {\bibfnamefont {R.}~\bibnamefont {Farrow}}, \bibinfo {author}
  {\bibfnamefont {L.~W.}\ \bibnamefont {Rupp}}, \bibinfo {author}
  {\bibfnamefont {A.~E.}\ \bibnamefont {White}}, \bibinfo {author}
  {\bibfnamefont {K.}~\bibnamefont {Short}}, \bibinfo {author} {\bibfnamefont
  {W.~F.}\ \bibnamefont {Peck}}, \ and\ \bibinfo {author} {\bibfnamefont
  {T.}~\bibnamefont {Kometani}},\ }\href {\doibase 10.1038/332814a0} {\bibfield
   {journal} {\bibinfo  {journal} {Nature}\ }\textbf {\bibinfo {volume}
  {332}},\ \bibinfo {pages} {814} (\bibinfo {year} {1988})}\BibitemShut
  {NoStop}%
\bibitem [{\citenamefont {Sleight}\ \emph {et~al.}(1975)\citenamefont
  {Sleight}, \citenamefont {Gillson},\ and\ \citenamefont
  {Bierstedt}}]{BaPbBiO3-Tc}%
  \BibitemOpen
  \bibfield  {author} {\bibinfo {author} {\bibfnamefont {A.}~\bibnamefont
  {Sleight}}, \bibinfo {author} {\bibfnamefont {J.}~\bibnamefont {Gillson}}, \
  and\ \bibinfo {author} {\bibfnamefont {P.}~\bibnamefont {Bierstedt}},\ }\href
  {\doibase 10.1016/0038-1098(75)90327-0} {\bibfield  {journal} {\bibinfo
  {journal} {Solid State Communications}\ }\textbf {\bibinfo {volume} {17}},\
  \bibinfo {pages} {27 } (\bibinfo {year} {1975})}\BibitemShut {NoStop}%
\bibitem [{\citenamefont {Nourafkan}\ \emph {et~al.}(2012)\citenamefont
  {Nourafkan}, \citenamefont {Marsiglio},\ and\ \citenamefont
  {Kotliar}}]{Kotliar-PRL}%
  \BibitemOpen
  \bibfield  {author} {\bibinfo {author} {\bibfnamefont {R.}~\bibnamefont
  {Nourafkan}}, \bibinfo {author} {\bibfnamefont {F.}~\bibnamefont
  {Marsiglio}}, \ and\ \bibinfo {author} {\bibfnamefont {G.}~\bibnamefont
  {Kotliar}},\ }\href {\doibase 10.1103/PhysRevLett.109.017001} {\bibfield
  {journal} {\bibinfo  {journal} {Phys. Rev. Lett.}\ }\textbf {\bibinfo
  {volume} {109}},\ \bibinfo {pages} {017001} (\bibinfo {year}
  {2012})}\BibitemShut {NoStop}%
\bibitem [{\citenamefont {Yin}\ \emph {et~al.}(2013)\citenamefont {Yin},
  \citenamefont {Kutepov},\ and\ \citenamefont {Kotliar}}]{Kotliar-PRX}%
  \BibitemOpen
  \bibfield  {author} {\bibinfo {author} {\bibfnamefont {Z.~P.}\ \bibnamefont
  {Yin}}, \bibinfo {author} {\bibfnamefont {A.}~\bibnamefont {Kutepov}}, \ and\
  \bibinfo {author} {\bibfnamefont {G.}~\bibnamefont {Kotliar}},\ }\href
  {\doibase 10.1103/PhysRevX.3.021011} {\bibfield  {journal} {\bibinfo
  {journal} {Phys. Rev. X}\ }\textbf {\bibinfo {volume} {3}},\ \bibinfo {pages}
  {021011} (\bibinfo {year} {2013})}\BibitemShut {NoStop}%
\bibitem [{\citenamefont {Bazhirov}\ \emph {et~al.}(2013)\citenamefont
  {Bazhirov}, \citenamefont {Coh}, \citenamefont {Louie},\ and\ \citenamefont
  {Cohen}}]{EPC-tilts}%
  \BibitemOpen
  \bibfield  {author} {\bibinfo {author} {\bibfnamefont {T.}~\bibnamefont
  {Bazhirov}}, \bibinfo {author} {\bibfnamefont {S.}~\bibnamefont {Coh}},
  \bibinfo {author} {\bibfnamefont {S.~G.}\ \bibnamefont {Louie}}, \ and\
  \bibinfo {author} {\bibfnamefont {M.~L.}\ \bibnamefont {Cohen}},\ }\href
  {\doibase 10.1103/PhysRevB.88.224509} {\bibfield  {journal} {\bibinfo
  {journal} {Phys. Rev. B}\ }\textbf {\bibinfo {volume} {88}},\ \bibinfo
  {pages} {224509} (\bibinfo {year} {2013})}\BibitemShut {NoStop}%
\bibitem [{\citenamefont {Bischofs}\ \emph {et~al.}(2002)\citenamefont
  {Bischofs}, \citenamefont {Kostur},\ and\ \citenamefont
  {Allen}}]{Allen-half-filling}%
  \BibitemOpen
  \bibfield  {author} {\bibinfo {author} {\bibfnamefont {I.~B.}\ \bibnamefont
  {Bischofs}}, \bibinfo {author} {\bibfnamefont {V.~N.}\ \bibnamefont
  {Kostur}}, \ and\ \bibinfo {author} {\bibfnamefont {P.~B.}\ \bibnamefont
  {Allen}},\ }\href {\doibase 10.1103/PhysRevB.65.115112} {\bibfield  {journal}
  {\bibinfo  {journal} {Phys. Rev. B}\ }\textbf {\bibinfo {volume} {65}},\
  \bibinfo {pages} {115112} (\bibinfo {year} {2002})}\BibitemShut {NoStop}%
\bibitem [{\citenamefont {Kostur}\ and\ \citenamefont
  {Allen}(1997)}]{Allen-single-carrier}%
  \BibitemOpen
  \bibfield  {author} {\bibinfo {author} {\bibfnamefont {V.~N.}\ \bibnamefont
  {Kostur}}\ and\ \bibinfo {author} {\bibfnamefont {P.~B.}\ \bibnamefont
  {Allen}},\ }\href {\doibase 10.1103/PhysRevB.56.3105} {\bibfield  {journal}
  {\bibinfo  {journal} {Phys. Rev. B}\ }\textbf {\bibinfo {volume} {56}},\
  \bibinfo {pages} {3105} (\bibinfo {year} {1997})}\BibitemShut {NoStop}%
\bibitem [{\citenamefont {Rice}\ and\ \citenamefont
  {Sneddon}(1981)}]{Rice-charge-disp}%
  \BibitemOpen
  \bibfield  {author} {\bibinfo {author} {\bibfnamefont {T.~M.}\ \bibnamefont
  {Rice}}\ and\ \bibinfo {author} {\bibfnamefont {L.}~\bibnamefont {Sneddon}},\
  }\href {\doibase 10.1103/PhysRevLett.47.689} {\bibfield  {journal} {\bibinfo
  {journal} {Phys. Rev. Lett.}\ }\textbf {\bibinfo {volume} {47}},\ \bibinfo
  {pages} {689} (\bibinfo {year} {1981})}\BibitemShut {NoStop}%
\bibitem [{\citenamefont {Cox}\ and\ \citenamefont
  {Sleight}(1976)}]{Cox-charge-disp-1}%
  \BibitemOpen
  \bibfield  {author} {\bibinfo {author} {\bibfnamefont {D.}~\bibnamefont
  {Cox}}\ and\ \bibinfo {author} {\bibfnamefont {A.}~\bibnamefont {Sleight}},\
  }\href {\doibase http://dx.doi.org/10.1016/0038-1098(76)90632-3} {\bibfield
  {journal} {\bibinfo  {journal} {Solid State Communications}\ }\textbf
  {\bibinfo {volume} {19}},\ \bibinfo {pages} {969 } (\bibinfo {year}
  {1976})}\BibitemShut {NoStop}%
\bibitem [{\citenamefont {Cox}\ and\ \citenamefont
  {Sleight}(1979)}]{Cox-charge-disp-2}%
  \BibitemOpen
  \bibfield  {author} {\bibinfo {author} {\bibfnamefont {D.~E.}\ \bibnamefont
  {Cox}}\ and\ \bibinfo {author} {\bibfnamefont {A.~W.}\ \bibnamefont
  {Sleight}},\ }\href {\doibase 10.1107/S0567740879002417} {\bibfield
  {journal} {\bibinfo  {journal} {Acta Crystallographica Section B}\ }\textbf
  {\bibinfo {volume} {35}},\ \bibinfo {pages} {1} (\bibinfo {year}
  {1979})}\BibitemShut {NoStop}%
\bibitem [{\citenamefont {Varma}(1988)}]{Varma-charge-disp}%
  \BibitemOpen
  \bibfield  {author} {\bibinfo {author} {\bibfnamefont {C.~M.}\ \bibnamefont
  {Varma}},\ }\href {\doibase 10.1103/PhysRevLett.61.2713} {\bibfield
  {journal} {\bibinfo  {journal} {Phys. Rev. Lett.}\ }\textbf {\bibinfo
  {volume} {61}},\ \bibinfo {pages} {2713} (\bibinfo {year}
  {1988})}\BibitemShut {NoStop}%
\bibitem [{\citenamefont {Hase}\ and\ \citenamefont
  {Yanagisawa}(2007)}]{Hase-charge-disp}%
  \BibitemOpen
  \bibfield  {author} {\bibinfo {author} {\bibfnamefont {I.}~\bibnamefont
  {Hase}}\ and\ \bibinfo {author} {\bibfnamefont {T.}~\bibnamefont
  {Yanagisawa}},\ }\href {\doibase 10.1103/PhysRevB.76.174103} {\bibfield
  {journal} {\bibinfo  {journal} {Phys. Rev. B}\ }\textbf {\bibinfo {volume}
  {76}},\ \bibinfo {pages} {174103} (\bibinfo {year} {2007})}\BibitemShut
  {NoStop}%
\bibitem [{\citenamefont {Sleight}(2015)}]{Sleight-review}%
  \BibitemOpen
  \bibfield  {author} {\bibinfo {author} {\bibfnamefont {A.~W.}\ \bibnamefont
  {Sleight}},\ }\href {\doibase 10.1016/j.physc.2015.02.012} {\bibfield
  {journal} {\bibinfo  {journal} {Physica C: Superconductivity and its
  Applications}\ }\textbf {\bibinfo {volume} {514}},\ \bibinfo {pages} {152 }
  (\bibinfo {year} {2015})},\ \bibinfo {note} {superconducting Materials:
  Conventional, Unconventional and Undetermined}\BibitemShut {NoStop}%
\bibitem [{\citenamefont {Nishio}\ \emph {et~al.}(2005)\citenamefont {Nishio},
  \citenamefont {Ahmad},\ and\ \citenamefont {Uwe}}]{Nishio}%
  \BibitemOpen
  \bibfield  {author} {\bibinfo {author} {\bibfnamefont {T.}~\bibnamefont
  {Nishio}}, \bibinfo {author} {\bibfnamefont {J.}~\bibnamefont {Ahmad}}, \
  and\ \bibinfo {author} {\bibfnamefont {H.}~\bibnamefont {Uwe}},\ }\href
  {\doibase 10.1103/PhysRevLett.95.176403} {\bibfield  {journal} {\bibinfo
  {journal} {Phys. Rev. Lett.}\ }\textbf {\bibinfo {volume} {95}},\ \bibinfo
  {pages} {176403} (\bibinfo {year} {2005})}\BibitemShut {NoStop}%
\bibitem [{\citenamefont {{Derimow}}\ \emph {et~al.}(2014)\citenamefont
  {{Derimow}}, \citenamefont {{Labry}}, \citenamefont {{Khodagulyan}},
  \citenamefont {{Wang}},\ and\ \citenamefont {{Zhao}}}]{Guo-meng-Zhao}%
  \BibitemOpen
  \bibfield  {author} {\bibinfo {author} {\bibfnamefont {N.}~\bibnamefont
  {{Derimow}}}, \bibinfo {author} {\bibfnamefont {J.}~\bibnamefont {{Labry}}},
  \bibinfo {author} {\bibfnamefont {A.}~\bibnamefont {{Khodagulyan}}}, \bibinfo
  {author} {\bibfnamefont {J.}~\bibnamefont {{Wang}}}, \ and\ \bibinfo {author}
  {\bibfnamefont {G.-m.}\ \bibnamefont {{Zhao}}},\ }\href@noop {} {\bibfield
  {journal} {\bibinfo  {journal} {ArXiv e-prints}\ } (\bibinfo {year}
  {2014})},\ \Eprint {http://arxiv.org/abs/1410.4100} {arXiv:1410.4100
  [cond-mat.supr-con]} \BibitemShut {NoStop}%
\bibitem [{\citenamefont {Menushenkov}\ and\ \citenamefont
  {Klementev}(2000)}]{Menushenkov-1}%
  \BibitemOpen
  \bibfield  {author} {\bibinfo {author} {\bibfnamefont {A.~P.}\ \bibnamefont
  {Menushenkov}}\ and\ \bibinfo {author} {\bibfnamefont {K.~V.}\ \bibnamefont
  {Klementev}},\ }\href {http://stacks.iop.org/0953-8984/12/i=16/a=303}
  {\bibfield  {journal} {\bibinfo  {journal} {Journal of Physics: Condensed
  Matter}\ }\textbf {\bibinfo {volume} {12}},\ \bibinfo {pages} {3767}
  (\bibinfo {year} {2000})}\BibitemShut {NoStop}%
\bibitem [{\citenamefont {Menushenkov}\ \emph {et~al.}(2001)\citenamefont
  {Menushenkov}, \citenamefont {Klementev}, \citenamefont {Kuznetsov},\ and\
  \citenamefont {Kagan}}]{Menushenkov-2}%
  \BibitemOpen
  \bibfield  {author} {\bibinfo {author} {\bibfnamefont {A.}~\bibnamefont
  {Menushenkov}}, \bibinfo {author} {\bibfnamefont {K.}~\bibnamefont
  {Klementev}}, \bibinfo {author} {\bibfnamefont {A.}~\bibnamefont
  {Kuznetsov}}, \ and\ \bibinfo {author} {\bibfnamefont {M.}~\bibnamefont
  {Kagan}},\ }\href {\doibase 10.1134/1.1410607} {\bibfield  {journal}
  {\bibinfo  {journal} {Journal of Experimental and Theoretical Physics}\
  }\textbf {\bibinfo {volume} {93}},\ \bibinfo {pages} {615} (\bibinfo {year}
  {2001})}\BibitemShut {NoStop}%
\bibitem [{\citenamefont {Mizokawa}\ \emph {et~al.}(2000)\citenamefont
  {Mizokawa}, \citenamefont {Khomskii},\ and\ \citenamefont
  {Sawatzky}}]{George-nickelates}%
  \BibitemOpen
  \bibfield  {author} {\bibinfo {author} {\bibfnamefont {T.}~\bibnamefont
  {Mizokawa}}, \bibinfo {author} {\bibfnamefont {D.~I.}\ \bibnamefont
  {Khomskii}}, \ and\ \bibinfo {author} {\bibfnamefont {G.~A.}\ \bibnamefont
  {Sawatzky}},\ }\href {\doibase 10.1103/PhysRevB.61.11263} {\bibfield
  {journal} {\bibinfo  {journal} {Phys. Rev. B}\ }\textbf {\bibinfo {volume}
  {61}},\ \bibinfo {pages} {11263} (\bibinfo {year} {2000})}\BibitemShut
  {NoStop}%
\bibitem [{\citenamefont {Park}\ \emph {et~al.}(2012)\citenamefont {Park},
  \citenamefont {Millis},\ and\ \citenamefont {Marianetti}}]{Millis1}%
  \BibitemOpen
  \bibfield  {author} {\bibinfo {author} {\bibfnamefont {H.}~\bibnamefont
  {Park}}, \bibinfo {author} {\bibfnamefont {A.~J.}\ \bibnamefont {Millis}}, \
  and\ \bibinfo {author} {\bibfnamefont {C.~A.}\ \bibnamefont {Marianetti}},\
  }\href {\doibase 10.1103/PhysRevLett.109.156402} {\bibfield  {journal}
  {\bibinfo  {journal} {Phys. Rev. Lett.}\ }\textbf {\bibinfo {volume} {109}},\
  \bibinfo {pages} {156402} (\bibinfo {year} {2012})}\BibitemShut {NoStop}%
\bibitem [{\citenamefont {Lau}\ and\ \citenamefont {Millis}(2013)}]{Millis2}%
  \BibitemOpen
  \bibfield  {author} {\bibinfo {author} {\bibfnamefont {B.}~\bibnamefont
  {Lau}}\ and\ \bibinfo {author} {\bibfnamefont {A.~J.}\ \bibnamefont
  {Millis}},\ }\href {\doibase 10.1103/PhysRevLett.110.126404} {\bibfield
  {journal} {\bibinfo  {journal} {Phys. Rev. Lett.}\ }\textbf {\bibinfo
  {volume} {110}},\ \bibinfo {pages} {126404} (\bibinfo {year}
  {2013})}\BibitemShut {NoStop}%
\bibitem [{\citenamefont {Adolphs}\ and\ \citenamefont
  {Berciu}(2013)}]{Clemens3}%
  \BibitemOpen
  \bibfield  {author} {\bibinfo {author} {\bibfnamefont {C.~P.~J.}\
  \bibnamefont {Adolphs}}\ and\ \bibinfo {author} {\bibfnamefont
  {M.}~\bibnamefont {Berciu}},\ }\href
  {http://stacks.iop.org/0295-5075/102/i=4/a=47003} {\bibfield  {journal}
  {\bibinfo  {journal} {EPL (Europhysics Letters)}\ }\textbf {\bibinfo {volume}
  {102}},\ \bibinfo {pages} {47003} (\bibinfo {year} {2013})}\BibitemShut
  {NoStop}%
\bibitem [{\citenamefont {Bon\ifmmode~\check{c}\else \v{c}\fi{}a}\ \emph
  {et~al.}(2008)\citenamefont {Bon\ifmmode~\check{c}\else \v{c}\fi{}a},
  \citenamefont {Maekawa}, \citenamefont {Tohyama},\ and\ \citenamefont
  {Prelov\ifmmode~\check{s}\else \v{s}\fi{}ek}}]{Bonca-t-J}%
  \BibitemOpen
  \bibfield  {author} {\bibinfo {author} {\bibfnamefont {J.}~\bibnamefont
  {Bon\ifmmode~\check{c}\else \v{c}\fi{}a}}, \bibinfo {author} {\bibfnamefont
  {S.}~\bibnamefont {Maekawa}}, \bibinfo {author} {\bibfnamefont
  {T.}~\bibnamefont {Tohyama}}, \ and\ \bibinfo {author} {\bibfnamefont
  {P.}~\bibnamefont {Prelov\ifmmode~\check{s}\else \v{s}\fi{}ek}},\ }\href
  {\doibase 10.1103/PhysRevB.77.054519} {\bibfield  {journal} {\bibinfo
  {journal} {Phys. Rev. B}\ }\textbf {\bibinfo {volume} {77}},\ \bibinfo
  {pages} {054519} (\bibinfo {year} {2008})}\BibitemShut {NoStop}%
\bibitem [{\citenamefont {Dagotto}(1994)}]{Dagotto-Lanczos}%
  \BibitemOpen
  \bibfield  {author} {\bibinfo {author} {\bibfnamefont {E.}~\bibnamefont
  {Dagotto}},\ }\href {\doibase 10.1103/RevModPhys.66.763} {\bibfield
  {journal} {\bibinfo  {journal} {Rev. Mod. Phys.}\ }\textbf {\bibinfo {volume}
  {66}},\ \bibinfo {pages} {763} (\bibinfo {year} {1994})}\BibitemShut
  {NoStop}%
\bibitem [{\citenamefont {Goodvin}\ \emph {et~al.}(2006)\citenamefont
  {Goodvin}, \citenamefont {Berciu},\ and\ \citenamefont
  {Sawatzky}}]{Mona-MA-3}%
  \BibitemOpen
  \bibfield  {author} {\bibinfo {author} {\bibfnamefont {G.~L.}\ \bibnamefont
  {Goodvin}}, \bibinfo {author} {\bibfnamefont {M.}~\bibnamefont {Berciu}}, \
  and\ \bibinfo {author} {\bibfnamefont {G.~A.}\ \bibnamefont {Sawatzky}},\
  }\href {\doibase 10.1103/PhysRevB.74.245104} {\bibfield  {journal} {\bibinfo
  {journal} {Phys. Rev. B}\ }\textbf {\bibinfo {volume} {74}},\ \bibinfo
  {pages} {245104} (\bibinfo {year} {2006})}\BibitemShut {NoStop}%
\end{thebibliography}

%merlin.mbs apsrev4-1.bst 2010-07-25 4.21a (PWD, AO, DPC) hacked
%Control: key (0)
%Control: author (8) initials jnrlst
%Control: editor formatted (1) identically to author
%Control: production of article title (-1) disabled
%Control: page (0) single
%Control: year (1) truncated
%Control: production of eprint (0) enabled
%

\end{document}